\documentclass[preprint,12pt]{elsarticle}

\usepackage{graphicx,import}
\usepackage{caption}
\usepackage{subcaption}
\usepackage{setspace}
\usepackage{xcolor}
\usepackage{siunitx} 
\usepackage{ulem} 

\usepackage{amssymb}
\usepackage{amsmath}
\usepackage[symbol]{footmisc}

\usepackage{lineno}
\usepackage{nomencl}
\usepackage{multicol}	
\makenomenclature
\usepackage{etoolbox}
\renewcommand\nomgroup[1]{%
  \item[\bfseries
  \ifstrequal{#1}{C}{Constants}{%
  \ifstrequal{#1}{G}{Greek letters}{%
  \ifstrequal{#1}{O}{Other symbols}{%
  \ifstrequal{#1}{S}{Subscripts}{}}}}%
]}

\journal{Elsevier}	

\begin{document}

\begin{spacing}{1.5}

\begin{frontmatter}

\title{Computational modelling of gas-liquid-solid multiphase free surface flow with and without evaporation}


\def\correspondingauthor{\footnote{Corresponding author: huihuang.xia@kit.edu}}
\author{Huihuang Xia\correspondingauthor, Marc Kamlah}

\address{Institute for Applied Materials (IAM), Karlsruhe Institute of Technology (KIT), Hermann-von Helmholtz-Platz 1 76344 Eggenstein-Leopoldshafen, Germany}

\begin{abstract}
Gas-liquid-solid multiphase systems are ubiquitous in engineering applications, e.g. inkjet printing, spray drying and coating. Developing a numerical framework for modelling these multiphase systems is of great significance. An improved, resolved CFD-DEM framework is developed to model the multiphase free surface flow with and without evaporation. An improved capillary force model is developed to compute the capillary interactions for partially floating particles at a free surface. Two well-known benchmark cases, namely drag coefficient calculation and the single sphere settling, are conducted to validate the resolved CFD-DEM model. It turns out that the resolved CFD-DEM model developed in this paper can accurately calculate the fluid-solid interactions and predict the trajectory of solid particles interacting with the liquid phase. Numerical demonstrations, namely two particles moving along a free surface when the liquid phase evaporates, and particle transport and accumulations inside an evaporating sessile droplet show the performance of the resolved model. 
\end{abstract}


\end{frontmatter}

\nomenclature[C]{$\boldsymbol{g}$}{Gravitational acceleration constant [\si{m/s^2}]}
\nomenclature[G]{$\alpha$}{Volume fraction [$-$]}
\nomenclature[G]{$\rho$}{Density [\si{kg/m^3}]}
\nomenclature[G]{$\tau$}{Viscous stress tensor}
\nomenclature[G]{$\theta$}{Contact angle}
\nomenclature[G]{$\Gamma$}{Boundary}
\nomenclature[G]{$\Sigma$}{Total stress tensor}
\nomenclature[G]{$\mu$}{Dynamic viscosity [\si{Pa \ s}]}
\nomenclature[O]{$\dot{m}$}{Mass source per unit volume [\si{kg/(m^3 \cdot s)}]}
\nomenclature[O]{$Y$}{Vapour mass fraction [$-$]}
\nomenclature[O]{$\epsilon_f$}{Void fraction [$-$]}
\nomenclature[O]{$\epsilon_s$}{Solid fraction [$-$]}
\nomenclature[O]{$p$}{Pressure [\si{Pa}]}
\nomenclature[O]{$D_v$}{Vapour diffusion coefficient [\si{m^2/s}]}
\nomenclature[O]{$\mathbf{t}_{\Gamma_s}$}{Traction vector}
\nomenclature[O]{$m$}{Mass}
\nomenclature[O]{$\mathbf{x}_i$}{Position vector}
\nomenclature[O]{$\mathbf{F}$}{Force}
\nomenclature[O]{$\mathbf{n}$}{Normal vector}
\nomenclature[O]{$\mathbf{U}$}{Velocity [\si{m/s}]}
\nomenclature[O]{$\mathbf{t}$}{Tangent vector}
\nomenclature[O]{$\mathbf{M}$}{Torque}
\nomenclature[O]{$\mathbf{I}$}{Identity tensor}
\nomenclature[S]{$l$}{Liquid}
\nomenclature[S]{$g$}{Gas}
\nomenclature[S]{$s$}{Solid}
\nomenclature[S]{$\text{fp}$}{Fluid-particle}
\nomenclature[S]{$\text{pf}$}{Particle-fluid}
\nomenclature[S]{$\text{cp}$}{Capillary}
\nomenclature[S]{$\text{st}$}{Surface tension}
\nomenclature[S]{$\text{evap}$}{Evaporation}
\nomenclature[S]{$i$}{The $i^{\text{th}}$ particle}
\nomenclature[S]{$ij$}{Particle $i$ - particle $j$}
\printnomenclature

\modulolinenumbers[2]	
\linenumbers

\section{Introduction}
\label{S:1}
Granular materials widely exist in our daily life (e.g. sugar, salt and coffee bins) and industry (e.g. coal, sand and bearing balls). However, some granular materials are surrounded by fluids such as gases or liquids (e.g. pneumatic conveying \cite{kuang2020cfd}, fluidized beds \cite{zhang2023numerical}, mudflow \cite{fang2022influence}, capillary suspensions \cite{koos2014capillary} and more \cite{golshan2020review, lu2022mfix, ma2022review}). Computational modelling of these kinds of gas-liquid-solid multiphase systems is of great importance for better understanding the complex interactions among solid particles and between the solid phase (particles) and the fluid phase (liquids or gases). 

The Discrete Element Method (DEM) is capable of modelling the complex mechanical behaviour of solid particles and the interactions between a particle and a wall. The motion of solid particles in different scales, namely macro-, meso- and micro-scale, can be tracked by solving Newton's second law of motion. Computational Fluid Dynamics (CFD) is generally used to model the motion of fluids by solving the Navier-Stokes equations. The so-called coupling approach, namely coupling CFD to DEM (CFD-DEM), is widely used to model the complex interaction between solid and fluid phases, as discussed below.

In this work, the resolved CFD-DEM approach is discussed. In the resolved CFD-DEM approach, the motions of solid and fluid phases are governed by DEM and CFD, respectively \cite{balachandran2021resolved}. Modelling enables us to understand multiphase systems at different scales, and extensive parameter studies can be conducted to investigate the influence of operating parameters at a lower cost. During the past decades, the CFD-DEM approach has been widely used to computationally model either compressible or incompressible particle-laden flow \cite{hager2014parallel, podlozhnyuk2018modelling, zhao2014investigation}. Hager et al. developed a simple method for smooth representation of the void fraction field for multi-scale resolved CFD-DEM simulations \cite{hager2014parallel, hager2014cfd}. Podlozhnyuk implemented the superquadric particles into the resolved CFD-DEM \cite{podlozhnyuk2018modelling}. Davydzenka et al. developed a resolved CFD-DEM model accounting for the wettability of complex geometry in multiphase flow \cite{davydzenka2020coupled}. Blood flow with irregular red blood cell particles was investigated within the resolved CFD-DEM framework by Balachandran et al. \cite{balachandran2021resolved}. Free surface flow with capillary interactions was studied by Nguyen et al., where a capillary force model was developed \cite{nguyen2021interface}. Flow with irregular particles constructed with multi-sphere clumps in an incompressible free surface flow was investigated by Shen et al. \cite{shen2022resolved}. Melting of solid particles for selective laser melting was computationally modelled by an improved resolved CFD-DEM approach developed by Yu et al. \cite{yu2021semi}. Schnorr Filho et al. investigated the hydraulic conveying of solid particles through a narrow elbow with a resolved CFD-DEM model \cite{schnorr2022resolved}. Free surface flow with superquadric particles was investigated by Washino et al., with the capillary force incorporated into the resolved CFD-DEM model \cite{washino2023development}. 

A summary of numerical simulations and applications of existing resolved CFD-DEM approaches is listed in Table~\ref{resolved_CFD-DEM_Summary}.
\begin{table}[h] 
\centering
\setlength\tabcolsep{2pt}
\footnotesize
\caption{Summary and comparison among these applications of resolved CFD-DEM formulations.}
\label{resolved_CFD-DEM_Summary}
\begin{tabular}{llcc}
\hline
Authors (publication year)  & Applications          & Surface tension & Phase change\\ \hline
Hager et al. (2014) \cite{hager2014parallel, hager2014cfd}       & Multi-scale modelling & no & no\\ 
Podlozhnyuk et al. (2017) \cite{podlozhnyuk2017efficient} & Flow with superquadric particles & no & no\\ 
Davydzenka et al. (2020) \cite{davydzenka2020coupled} & Multiphase flow in porous media & yes & no\\
Balachandran Nair et al. (2021) \cite{balachandran2021resolved} & Blood flow in microfluidic devices & no & no\\ 
Nguyen et al. (2021) \cite{nguyen2021interface} & Flow with capillary interactions & yes & no\\
Shen et al. (2022) \cite{shen2022resolved} & Flow with irregular particles & yes & no\\ 
Yu et al. (2021) \cite{yu2021semi} & Selective laser melting & yes & yes\\ 
Schnorr Filho et al. (2022) \cite{schnorr2022resolved} & Hydraulic conveying & no & no\\ 
Washino et al. (2023) \cite{washino2023development} & Flow with non-spherical particles & yes & no\\ \hline
\end{tabular}
\end{table}
However, what can be seen from the table is that a limited number of publications can be found related to model solid particles immersed in an incompressible flow that undergoes phase change or evaporation. Direct inkjet printing \cite{derby2015additive, lohse2022fundamental}, spray drying \cite{giuliano2023micromechanics, yang2024micro} and spray coating \cite{kieckhefen2019simulation, christodoulou2020model} involve complex solid-liquid interactions, the phase change from liquid to vapour, surface tension and beyond. Accordingly, developing such a numerical model accounting for additional phase change and surface tension is of great significance. 

In this paper, an improved resolved CFD-DEM framework incorporating free surface capturing, surface tension and phase change of the liquid phase is developed by extending the resolved CFD-DEM model developed by Hager et al. \cite{hager2014parallel, hager2014cfd}. The new resolved CFD-DEM framework is implemented in the open-source framework CFDEMcoupling-PUBLIC \cite{githubDCS} bridging the open-source DEM code LIGGGHTS \cite{githubLIGGGHTS} and the open-source Finite Volume Method (FVM) based C++ library OpenFOAM \cite{OpenFOAM5X}. The large-scale parallel computation and data exchange between the two codes are realized using the Message Passing Interface (MPI) software \cite{gabriel2004open}.

This paper consists of the following sections: the mathematical formulation, namely the governing equations for the liquid and solid phases, are presented in Section~\ref{MathFormu}. The detailed numerical method for solving these governing equations, numerical procedure and coupling algorithm are discussed in Section~\ref{NumMethod}. In Section~\ref{Results}, the numerical validations and demonstrations are presented. The conclusions of this paper are summarized in Section~\ref{Conclusion}. Some additional contents are discussed in the Appendix of this paper for completeness.

\section{Mathematical formulation}
\label{MathFormu}
In this section, the theory and mathematical formulation of the resolved CFD-DEM approach are introduced in detail. 
\begin{figure}[h]
  \begin{center}
    \includegraphics[width=1.0\textwidth]{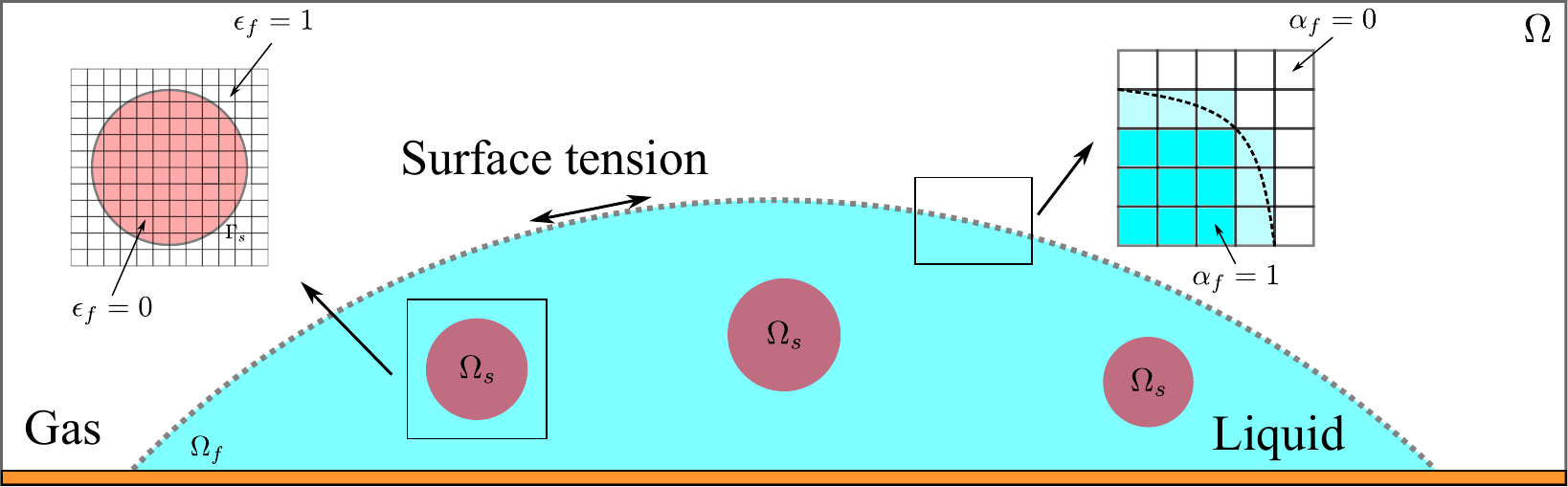}
  \end{center}
  \caption{The schematic diagram of a surface-tension-dominant gas-liquid-solid multiphase system.}
  \label{diagram_ResolvedCFD-DEM}
\end{figure}
For the resolved CFD-DEM approach, one solid particle usually occupies several CFD cells\footnote{Here, the CFD cell refers to a small computational cell bounded by arbitrary polygonal faces after discretizing the computational domain, numerically.} (see the inset on the top-left of Figure~\ref{diagram_ResolvedCFD-DEM}). The void fraction field $\epsilon_f$ is used to quantify how much volume is not occupied by a solid particle in each CFD cell, because of which $\epsilon_f = 0$ indicates that a solid particle fully covers the current CFD cell. The free surface is captured by the VoF method incorporated in the i-CLSVoF framework presented in our previous work \cite{xia2022improved}. The Fictitious Domain Method developed by Patankar et al. \cite{patankar2000new} was extended to incorporate the free-surface capturing, surface tension and evaporation for modelling the complex multiphase system as demonstrated by Figure~\ref{diagram_ResolvedCFD-DEM}.

\begin{figure}[h]
 \begin{center}
    \includegraphics[width=0.5\textwidth]{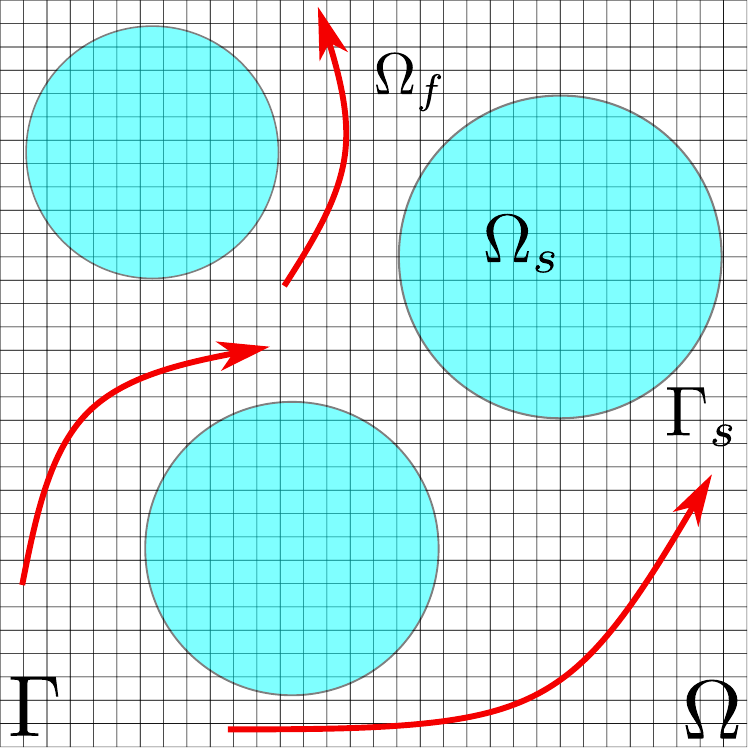}
  \end{center}
  \caption{The schematic diagram of the resolved CFD-DEM model. Arrows represent the vectors of fluid velocity around particles, and circles in blue represent solid particles.}
  \label{ResolvedCFD-DEM}
\end{figure}

As shown in Figure~\ref{ResolvedCFD-DEM}, $\Omega$ is the whole computational domain, and $\Omega_f$ and $\Omega_s$ are the liquid and solid phases, respectively. $\Gamma$ and $\Gamma_s$ denote the boundaries of the whole computational domain and the solid particles immersed in the liquid, respectively. 
The governing equations for the surface-tension-dominant incompressible Newtonian flow are given by
\begin{equation} \label{ConEqn-ReCFDEM}
\nabla \cdot{\mathbf{U}}=\epsilon_f \dot{m} (\frac {1}{\rho_g}-\frac {1}{\rho_l}) \ \text{in} \ \Omega,
\end{equation}
\begin{equation} \label{MomEqn-ReCFDEM}
\frac{\partial(\rho\mathbf{U})}{\partial t}+\nabla \cdot (\rho\mathbf{U}\mathbf{U})=-\nabla p+\nabla \cdot \left[\mu \left(\nabla \mathbf{U} + (\nabla \mathbf{U})^{\text{T}} \right) \right]+\rho \mathbf{g}+ \mathbf{F}_{\text{st}} \  \text{in} \ \Omega,
\end{equation}
where $\epsilon_f$ is the void fraction field. The incorporation of $\epsilon_f$ is to guarantee that phase change does not occur inside solid particles. $\dot{m}$ in the continuity equation (Eqn.~\ref{ConEqn-ReCFDEM}) is the mass source term per unit volume, where $\dot{m}=0$ holds for the case without evaporation or phase change only. Furthermore, some other terms, e.g. pressure gradient $\nabla p$ and surface tension force $\mathbf{F}_{\text{st}}$ in the momentum equation (Eqn.~\ref{MomEqn-ReCFDEM}) are detailed in our previous work \cite{xia2022improved}.

The governing equations and numerical method for the computational modelling of droplet evaporation are detailed in our previous work \cite{xia2022improved}. This section only mentions some basic equations for modelling evaporation. The vapour mass fraction gradient model discussed in our previous work is used in this paper. The vapour mass fraction $Y$ is solved from
\begin{equation} \label{YEqn-Resolved_CFD-DEM}
\frac{\partial Y}{\partial t}+\nabla \cdot (Y \mathbf{U})=D_v \nabla^{2} Y,
\end{equation}
where $D_v$ is the vapour diffusion coefficient \cite{xia2022improved}. The mass source term $\dot m$ in the continuity equation (Eqn.~\ref{ConEqn-ReCFDEM}) is calculated by
\begin{equation} \label{massFluxVapour-dotM}
\dot m = \frac{\rho_g D_v \nabla^{\Gamma} Y \mathbf{n}^{\Gamma}}{1-Y^{\Gamma}} \lvert \nabla \alpha_l \rvert,
\end{equation}
where $\alpha_l$ is the liquid volume fraction field. Some other quantities (e.g. $\mathbf{n}^{\Gamma}$ and $Y ^{\Gamma}$) and the numerical method for the evaporation model are detailed in our previous work \cite{xia2022improved}. 

Newton's second law of motion governs the motion of solid particles. The equation for the translational motion is given by
\begin{equation} \label{DEMGEqn1-ReCFDEM}
m_i\frac{\partial^{2}\mathbf{x}_i}{\partial t^{2}}=\sum_{i=1}^{N_p}\mathbf{F}_{ij}+m_i\mathbf{g}+\mathbf{F}_{\text{fp}}^\text{c}+\mathbf{F}_{\text{cp}},
\end{equation}
where $m_i$ and $\mathbf{x}_i$ are the mass and position vector of the $i^{\text{th}}$ particle, respectively, $\mathbf{F}_{ij}$ is the contact force between two DEM elements (particle-particle or particle-wall), and $\mathbf{F}_{\text{fp}}^\text{c}$ the CFD-DEM coupling force acting on the solid particles by the liquid phase. The last term on the right-hand side of Eqn.~\ref{DEMGEqn1-ReCFDEM} is the capillary force acting on solid particles, which is discussed in the next section in detail. Besides, the rotational motion of solid particles is governed by
\begin{equation} \label{DEMGEqn2-ReCFDEM}
I_i\frac{\partial^{2}\mathbf{\theta}_i}{\partial t^{2}} = \sum_{i=1}^{N_p}\mathbf{M}_{ij} + \mathbf{M}_{\text{fp}}^\text{c}+\mathbf{M}_{\text{cp}},
\end{equation} 
where $I_i$ and $\mathbf{\theta}_i$ are the moment of inertia and angular displacement of the particle $i$, respectively. $\mathbf{M}_{ij}$ is the torque acting on particle $i$ by some other particles interacting with it. $\mathbf{M}_{\text{fp}}^\text{c}$ is the coupling term accounting for the torque acting on the solid phase by the liquid phase. The last term $\mathbf{M}_{\text{cp}}$ on the right-hand side of Eqn.~\ref{DEMGEqn2-ReCFDEM} is the torque acting on the particle $i$ due to capillary interactions. Calculations of these terms are introduced in the forthcoming sections.

\section{Numerical method}
\label{NumMethod}
Accurate calculations of interaction forces acting on the solid phase by the liquid phase are of great importance for realizing the resolved CFD-DEM. 

\subsection{Calculations of the interaction forces and torque}
In the Fictitious Domain Method, additional boundary and interface conditions are needed to be applied due to the presence of solid particles in the liquid phase \cite{shirgaonkar2009new}. These additional conditions are given by
\begin{equation} \label{additionalBCs}
\begin{cases}
\mathbf{U} = \mathbf{U}_{\Gamma} \ \text{on} \ \Gamma,\\
\mathbf{U} = \mathbf{U}_i \ \text{on} \ \Omega_s,\\
\boldsymbol{\Sigma} \cdot \mathbf{n} = \mathbf{t}_{\Gamma_s} \ \text{on} \ \Gamma_s,\\
\mathbf{U} (\mathbf{x}, \ t=0) = \mathbf{U}_0 (\mathbf{x}) \ \text{in} \ \Omega_f.\\
\end{cases}
\end{equation}

Here, the subscript $i$ indicates the $i^{\text{th}}$ particle and $\boldsymbol{\Sigma}$ is the total stress tensor as discussed below, and $\mathbf{n}$ is the outward normal vector to $\Gamma_s$. $\mathbf{t}_{\Gamma_s}$ is the traction vector acting from the liquid phase on the surface of solid particles. The second and third equations of Eqn.~\ref{additionalBCs} are responsible for the coupling between liquid and solid phases. Additionally, the second equation ensures the transfer of the particle velocity $\mathbf{U}_i$ to the liquid velocity of CFD cells covered by the solid particle $i$. The third equation represents the force acting on the boundaries of the solid phase.

The force acting on the solid phase by the liquid phase can be calculated by integrating the third equation of Eqn.~\ref{additionalBCs} over the whole solid surface as
\begin{equation} \label{F_Drag}
\mathbf{F}_{\text{fp}}^\text{c} = \int_{\Gamma_s} \mathbf{t}_{\Gamma_s} \,dS.
\end{equation}
The surface integral can be transformed to a volume integral using the divergence theorem which leads to
\begin{equation} \label{DivThem}
\int_{\Gamma_s} \mathbf{t}_{\Gamma_s} \ dS = \int_{\Gamma_s} \boldsymbol{\Sigma} \cdot \mathbf{n} \ dS =  \int_{\Omega_s} \nabla \cdot \boldsymbol{\Sigma} \,dV.
\end{equation}
The total stress tensor $\boldsymbol{\Sigma}$ in Eqn.~\ref{DivThem} consists of two terms given by
\begin{equation} \label{totalStressT}
\boldsymbol{\Sigma} = - p \mathbf{I} + \mathbf{\tau},
\end{equation}
where $p$ is the pressure and $\mathbf{I}$ the identity tensor of size $3 \times 3$, while $\mathbf{\tau}$ is the viscous stress tensor. For incompressible Newtonian fluids, $\mathbf{\tau}$ is deviatoric and given by
\begin{equation} \label{visStressT}
\mathbf{\tau} = \mu \left(\nabla \mathbf{U}+(\nabla \mathbf{U} \right)^\text{T}).
\end{equation}

Substituting Eqs.~\ref{DivThem}, \ref{totalStressT} and \ref {visStressT} into Eqn.~\ref{F_Drag}, the interaction force acting on the solid phase can be calculated by 
\begin{equation} \label{F_DragCal}
\begin{split}
\mathbf{F}_{\text{fp}}^\text{c} & = \int_{\Omega_s} \nabla \cdot \boldsymbol{\Sigma} \,dV\\
& =\int_{\Omega_s} \nabla \cdot \left[-p \mathbf{I}+\mu \left(\nabla \mathbf{U}+(\nabla \mathbf{U})^\text{T} \right) \right] \,dV\\
& = \int_{\Omega_s} \underbrace{\nabla \cdot (-p \mathbf{I})}_{= - \nabla p} + \nabla \cdot \left[\mu \left(\nabla \mathbf{U}+(\nabla \mathbf{U} \right)^\text{T}) \right] \,dV.
\end{split}
\end{equation}

As derived in the literature \cite{rusche2003computational}, for incompressible Newtonian fluids, the divergence of the viscous stress tensor is given by
\begin{equation} \label{visStressTensor}
\nabla \cdot \mathbf{\tau} = \nabla \cdot \left[\mu \left(\nabla \mathbf{U}+(\nabla \mathbf{U} \right)^\text{T}) \right] = \nabla \cdot (\mu \nabla \mathbf{U}) + \nabla \mathbf{U} \cdot \nabla \mu.
\end{equation}

In this work, one assumption is that the dynamic viscosity $\mu$ defined by the one-field formulation ($\mu = \alpha_1 \mu_1 + (1-\alpha_1) \mu_2$ with $\alpha_1$ and $\mu_1$ being the volume fraction and dynamic viscosity of phase 1, respectively) is constant, and thus Eqn.~\ref{visStressTensor} leads to 
\begin{equation} \label{visStressTensor_assumption}
\nabla \cdot \mathbf{\tau} = \nabla \cdot \left[\mu \left(\nabla \mathbf{U}+(\nabla \mathbf{U} \right)^\text{T}) \right] = \mu \nabla^2 \mathbf{U}.
\end{equation}

Substituting Eqn.~\ref{visStressTensor_assumption} into Eqn.~\ref{F_DragCal}, leads to 
\begin{equation} \label{F_DragCal2}
\begin{split}
\mathbf{F}_{\text{fp}}^\text{c} = \int_{\Omega_s} \nabla \cdot \left[-p \mathbf{I}+\mu \left(\nabla \mathbf{U}+(\nabla \mathbf{U})^\text{T} \right) \right] \,dV = \int_{\Omega_s} \left(- \nabla p + \mu \nabla^2 \mathbf{U} \right) \,dV.
\end{split}
\end{equation}

The volume integral in Eqn.~\ref{F_DragCal2} can be approximated  by summarizing $- \nabla p + \mu \nabla^2 \mathbf{U}$ at cell centers of all the cells either partially or fully covered by a solid particle as 
\begin{equation}
\mathbf{F}_{\text{fp}}^\text{c} \approx \sum_{i=1}^{N_c}(- \nabla p + \mu \nabla^2 \mathbf{U})V_c,
\end{equation}
where $N_c$ is the total number of cells, and $V_c$ is the cell volume.

Similarly, the torque $\mathbf{M}_{\text{fp}}^\text{c}$ acting on particles by the liquid phase can be calculated by
\begin{equation}
\mathbf{M}_{\text{fp}}^\text{c} = \int_{\Gamma_s} \mathbf{r} \times \mathbf{t}_{\Gamma_s} \,dS = \int_{\Omega_s} \mathbf{r} \times (-\nabla p +\nabla \cdot \mathbf{\tau}) \,dV,
\end{equation}
where $\mathbf{r}$ is the position vector. $\mathbf{M}_{\text{fp}}^\text{c}$ is approximated by summarizing all these quantities as
\begin{equation}
\mathbf{M}_{\text{fp}}^\text{c} \approx \sum_{i=1}^{N_c} \mathbf{r} \times \left(- \nabla p + \mu \nabla^2 \mathbf{U} \right)V_c.
\end{equation} 
The last terms on the right-hand side of Eqs.~\ref{DEMGEqn1-ReCFDEM} and \ref{DEMGEqn2-ReCFDEM} are due to the capillary interactions among particles protruding from the free surface of liquids. The capillary force is of great importance for surface-tension-dominant flow because capillary interactions govern the motions, leading to self-assembly or self-organization of particles which appear at a free surface \cite{fujita2013computation}. The capillary force is a long-range attractive force which is more dominant than other forces, e.g. inertial force and gravitational force for surface-tension-dominant cases \cite{uzi2016modeling}. Thus, this force is accounted for in this work. 

\begin{figure}[h]
 \begin{center}
    \includegraphics[width=0.75\textwidth]{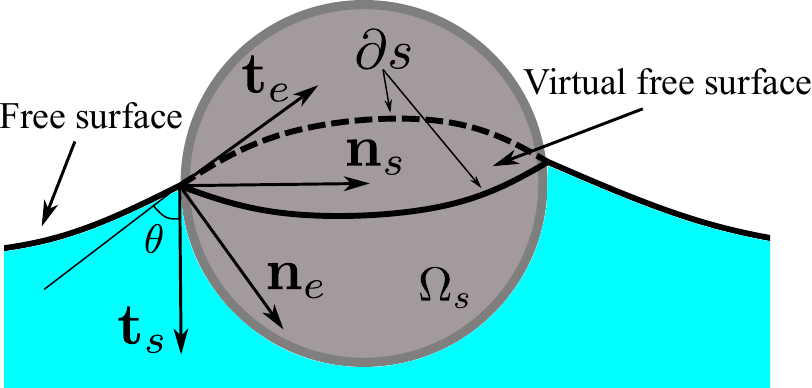}
  \end{center}
  \caption{The schematic diagram of the capillary force model. The virtual free surface is bounded by the solid and dashed three-phase contact lines $\partial s$.}
  \label{capillaryForce}
\end{figure}

As proven by Fujita et al. in the literature \cite{fujita2013computation}, the sum of the surface tension force along the three-phase contact line $\partial s$ is equal to the sum of the surface tension force over the virtual free surface fully immersed inside the solid particle in three dimensions as shown in Figure~\ref{capillaryForce}. 

The Immersed Free Surface model proposed in the literature \cite{nguyen2021interface, washino2023development, fujita2015direct} is extended to account for the wettability of solid particles at a free surface in this work. The basic idea is to solve the extrapolation equation
\begin{equation} \label{extraEqn-modelA}
\frac{\partial \alpha_l}{\partial t} + \mathbf{t}_e \cdot \nabla \alpha_l = 0
\end{equation}
in the true particle domain, namely, a domain with $\epsilon_f < 0.5$ to extrapolate the free surface from the liquid phase inside solid particles \cite{nguyen2021interface}. In Eqn.~\ref{extraEqn-modelA}, $\mathbf{t}_e$ is the tangent vector to the liquid surface pointing towards the particle. It is used to extend the liquid volume fraction and is defined by
\begin{equation}
\mathbf{t}_e = \frac{\mathbf{n}_s - (\mathbf{n}_e \cdot \mathbf{n}_s)\mathbf{n}_e}{|\mathbf{n}_s - (\mathbf{n}_e \cdot \mathbf{n}_s)\mathbf{n}_e|}
\end{equation}
with $\mathbf{n}_e$ being the normal vector to the liquid pointing inside the liquid, which is given by
\begin{equation}
\mathbf{n}_e = \mathbf{n}_s \text{cos} \theta + \mathbf{t}_s \text{sin} \theta.
\end{equation}
Here $\theta$ is the contact angle between the surface of the solid particle and the free surface. The normal vector to the particle surface pointing inwards $\mathbf{n}_s$ and  the unit vector $\mathbf{t}_s$ perpendicular to $\mathbf{n}_s$ are defined by
\begin{equation}
\mathbf{n}_s = \frac{\nabla \epsilon_s}{|\nabla \epsilon_s|}
\end{equation}
and
\begin{equation}
\mathbf{t}_s = \frac{\nabla \alpha_l - (\mathbf{n}_s \cdot \nabla \alpha_l)\mathbf{n}_s}{|\nabla \alpha_l - (\mathbf{n}_s \cdot \nabla \alpha_l)\mathbf{n}_s|},
\end{equation}
respectively, where $\epsilon_s$ is the solid fraction defined by $1.0 - \epsilon_f$.

In addition to the extrapolation equation given by Eqn.~\ref{extraEqn-modelA}, another model proposed by Fujita et al. \cite{fujita2015direct} is also implemented in this work. The first model given by Eqn.~\ref{extraEqn-modelA} is named as \textit{Model A} and the other model is called \textit{Model B}. The extrapolation equation of \textit{Model B} is given by
\begin{equation} \label{extraEqn-modelB}
\frac{\partial \alpha_l}{\partial t} + \epsilon_s \mathbf{n}_s \cdot \nabla \alpha_l = \epsilon_s |\nabla \alpha_l| \text{cos} \theta.
\end{equation}

Solving a diffusion equation and then an anti-diffusion equation to smoothen the liquid volume fraction field $\alpha_l$ and simultaneously suppressing the interface diffusion after solving Eqn.~\ref{extraEqn-modelA} or Eqn.~\ref{extraEqn-modelB} to guarantee more numerical stability was proposed by Nguyen et al. \cite{nguyen2021interface}. In this work, a simple approach without interface diffusion is applied, namely, the Laplacian filter approach proposed by Lafaurie et al. \cite{lafaurie1994modelling} is adopted to transform $\mathbf{t}_e \cdot \nabla \alpha_l$ in Eqn.~\ref{extraEqn-modelA} or $\mathbf{n}_s \cdot \nabla \alpha_l$ in Eqn.~\ref{extraEqn-modelB} into a smoother function $\widetilde{\mathbf{t}_e \cdot \nabla \alpha_l}$ or $\widetilde{\mathbf{n}_s \cdot \nabla \alpha_l}$. The transformations for $\mathbf{t}_e \cdot \nabla \alpha_l$ and $\mathbf{n}_s \cdot \nabla \alpha_l$ are given by
\begin{equation} \label{smoothFuncA}
\widetilde{\mathbf{t}_e \cdot \nabla \alpha_l} = \frac{\sum_{f=1}^n (\mathbf{t}_e \cdot \nabla \alpha_l)_f S_f}{\sum_{f=1}^n S_f},
\end{equation}
\begin{equation} \label{smoothFuncB}
\widetilde{\mathbf{n}_s \cdot \nabla \alpha_l} = \frac{\sum_{f=1}^n (\mathbf{n}_s \cdot \nabla \alpha_l)_f S_f}{\sum_{f=1}^n S_f},
\end{equation}
respectively, where $S_f$ is the magnitude of the $f^{\text{th}}$ face area of the computational cell which is bounded by $n$ faces, and $f$ denotes the face index. The value of $(\mathbf{t}_e \cdot \nabla \alpha_l)_f$ or $(\mathbf{n}_s \cdot \nabla \alpha_l)_f$ at the face center is calculated using linear interpolation over the interface region. The smooth function is then used to solve either Eqn.~\ref{extraEqn-modelA} or Eqn.~\ref{extraEqn-modelB} to construct a smooth virtual free surface. An artificial correction of the liquid volume fraction field $\alpha_l$ given by
\begin{equation} \label{boundedness}
\alpha_l = \text{max} \bigl( 0, \text{min}(1, \alpha_l) \bigl)
\end{equation}
after solving Eqn.~\ref{extraEqn-modelA} or Eqn.~\ref{extraEqn-modelB} is conducted to guarantee the boundedness of $\alpha_l$. Artificial boundedness of the liquid volume fraction field inside the solid phase by Eqn.~\ref{boundedness} leads to mass conservation issues, namely, the total liquid volume in the whole computational domain is not conserved. A numerical correction step is then adopted to correct the liquid volume fraction in the liquid phase. First, the total liquid volume of the pure liquid phase before solving Eqn.~\ref{extraEqn-modelA} or Eqn.~\ref{extraEqn-modelB} denoted as $V_{\text{before}}$ is calculated by
\begin{equation}
V_{\text{before}} = \int_{\Omega} \epsilon_f \alpha_l \,dV.
\end{equation}
Second, the total liquid volume of the pure liquid phase after solving Eqn.~\ref{extraEqn-modelA} or Eqn.~\ref{extraEqn-modelB} denoted as $V_{\text{after}}$ is calculated by
\begin{equation}
V_{\text{after}} = \int_{\Omega} \epsilon_f \alpha_l \,dV.
\end{equation}
The volume loss of the pure liquid phase due to the evaporation $V_{\text{evap}}$ is calculated by
\begin{equation}
V_{\text{evap}} = \int_{\Omega} \frac{\dot{m}}{\rho_l}  \Delta t \,dV,
\end{equation}
where $\Delta t$ is the time step, and $\rho_l$ is the density of the liquid phase. Accordingly, the volume change $\Delta V$ is computed by
\begin{equation}
\Delta V = V_{\text{after}} - V_{\text{before}} - V_{\text{evap}}.
\end{equation}
The liquid volume fraction field $\alpha_l$ in the liquid phase ($\epsilon_f > 0.5$) can be corrected by
\begin{equation} \label{numCorrect}
\alpha_l = \alpha_l + \frac{\epsilon_f \Delta V |\nabla \alpha_l|}{S_s}.
\end{equation}
Here $S_s$ is the total surface area of the pure liquid phase calculated by
\begin{equation}
S_s = \int_{\Omega} \epsilon_f |\nabla \alpha_l| \,dV.
\end{equation}

The numerical procedure to construct the virtual free surface inside solid particles using the Immersed Free Surface model is outlined as follows:
\begin{itemize}
\item{Smoothening either $\mathbf{t}_e \cdot \nabla \alpha_l$ with Eqn.~\ref{smoothFuncA} or $\mathbf{n}_s \cdot \nabla \alpha_l$ with Eqn.~\ref{smoothFuncB} before solving Eqn.~\ref{extraEqn-modelA} for \textit{Model A} and Eqn.~\ref{extraEqn-modelB} for \textit{Model B}, respectively.}
\item{Solving either Eqn.~\ref{extraEqn-modelA} or Eqn.~\ref{extraEqn-modelB} to construct a virtual free surface inside solid particles.}
\item{Using Eqn.~\ref{boundedness} to correct and constrain the liquid volume fraction field $\alpha_l$, artificially.}
\item{Correcting the liquid volume fraction in the liquid phase with Eqn.~\ref{numCorrect}.}
\end{itemize}

\begin{figure}[h]
 \begin{center}
    \includegraphics[width=0.5\textwidth]{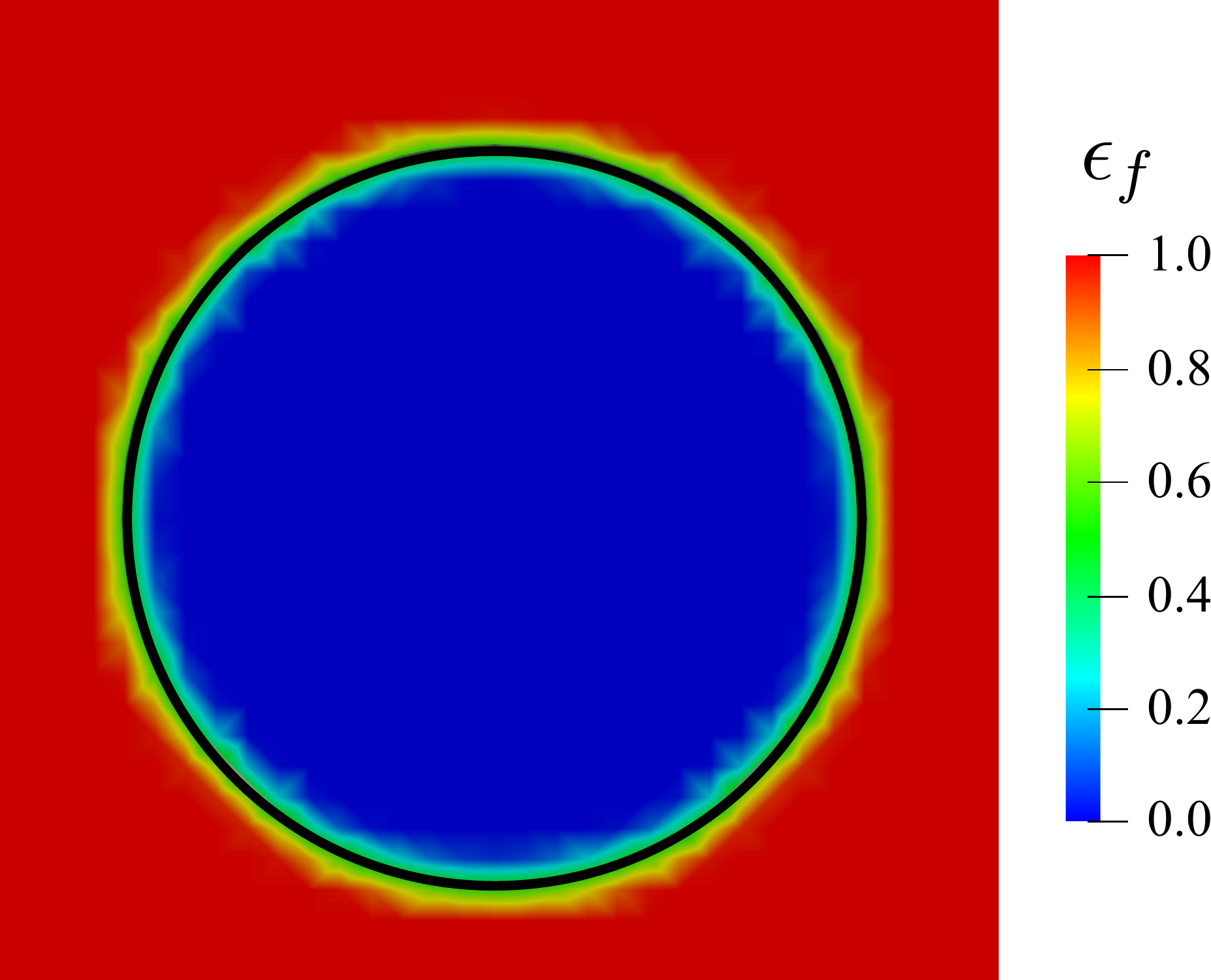}
  \end{center}
  \caption{The void fraction field of a solid particle and the solid black curve represents the iso-surface with $\epsilon_f = 0.5$.}
  \label{voidFractionField}
\end{figure}

Once constructing the virtual free surface inside solid particles as shown in Figure~\ref{capillaryForce}, the capillary force can be calculated by
\begin{equation} \label{lineIntegral}
\mathbf{F}_{\text{cp}} = \oint_{\partial s} \sigma K \mathbf{n} \,ds,
\end{equation}
where $ds$ is the infinitesimal integration area, and $\sigma$ and $K$ are the surface tension coefficient and mean interface curvature, respectively. As derived in Appendix B of \cite{tryggvason2011direct}, the surface integral over the immersed free surface $S$ in Eqn.~\ref{lineIntegral} can be transformed into a volume integral enclosing the immersed free surface $S$ shown in Figure~\ref{capillaryForce} given by
\begin{equation} 
\mathbf{F}_{\text{cp}} = \int_{\Omega_s} \sigma K \mathbf{n} \delta_s \,dV,
\end{equation}
where $\delta_s$ is a Dirac function \cite{tryggvason2011direct}. The term $\sigma K \mathbf{n} \delta_s$ can be can be summarized as the surface tension force $\mathbf{F}_{\text{st}}$. As suggested in the literature \cite{nguyen2021interface}, the capillary force can then be calculated over the true particle domain, namely domain with $\epsilon_f < 0.5$ (see Figure~\ref{voidFractionField}) by
\begin{equation}  \label{capForce}
\mathbf{F}_{\text{cp}} = \int_{{\Omega_s}^{\epsilon_f < 0.5}} \mathbf{F}_{\text{st}} \,dV.
\end{equation}
Similarly, the torque resulting from the capillary force can be calculated by
\begin{equation} \label{torque_cp}
\mathbf{M}_{\text{cp}} = \int_{{\Omega_s}^{\epsilon_f < 0.5}} \mathbf{r} \times \mathbf{F}_{\text{st}} \,dV,
\end{equation}
once the surface tension force $\mathbf{F}_{\text{st}}$ is known.

However, some preliminary numerical simulations demonstrate that this approach suffers from un-physical spurious velocities inside solid particles, which is also discussed in the literature \cite{washino2023development}. Therefore, the filtered surface tension force model implemented and discussed in our previous work \cite{xia2022improved} is adopted here to calculate the capillary force and the torque as
\begin{equation}
\mathbf{F}_{\text{cp}} = \int_{{\Omega_s}^{\epsilon_f < 0.5}} \mathbf{F}_{\text{st},f}^f \,dV
\end{equation}
and 
\begin{equation}
\mathbf{M}_{\text{cp}} = \int_{{\Omega_s}^{\epsilon_f < 0.5}} \mathbf{r} \times \mathbf{F}_{\text{st},f}^f \,dV,
\end{equation}
respectively, where $\mathbf{F}_{\text{st},f}^f$ is the filtered surface tension force.

Alternatively, the term $\mathbf{F}_{\text{st}}$ in Eqs.~\ref{capForce} and \ref{torque_cp} can be replaced by $\mathbf{F}_{\text{ccf}}$ calculated by the Continuous Capillary Force (CCF) model \cite{washino2013new} which is given by
\begin{equation}
\mathbf{F}_{\text{ccf}} = \sigma \mathbf{t}_c (\nabla \alpha_l \cdot \mathbf{t}_s)(\nabla \epsilon_s \cdot \mathbf{n}_s),
\end{equation}
where $\mathbf{t}_c$ is given by
\begin{equation}
\mathbf{t}_c = -\frac{\mathbf{n}_s - (\mathbf{n}_c \cdot \mathbf{n}_s)\mathbf{n}_c}{|\mathbf{n}_s - (\mathbf{n}_c \cdot \mathbf{n}_s)\mathbf{n}_c|}.
\end{equation}

In the resolved CFD-DEM approach, one solid particle covers several CFD cells, and thus the void fraction $\epsilon_f$ is of great importance in calculating $\mathbf{F}_{\text{fp}}^\text{c}$, $\mathbf{M}_{\text{fp}}^\text{c}$ and some other quantities, accurately. The smooth representation algorithm proposed by Hager \cite{hager2014cfd} is used in this paper to create a smooth transition of the void fraction around the particle surface. It is proven that this algorithm is more stable than the conventional stair-step representation algorithm \cite{hager2014cfd} and guarantees reasonable numerical accuracy. 

\subsection{The numerical procedure}
\label{NumCorrections}
In literature, an additional force term is incorporated to the right-hand side of the momentum Eqn.~\ref{MomEqn-ReCFDEM} to account for the interaction force acting on the fluid phase by the solid phase, which is known as the direct forcing approach  \cite{balachandran2021resolved, washino2023development, uhlmann2005immersed, wu2020forcing}. Instead of this, the numerical correction approach proposed in the literature as well \cite{hager2014parallel, podlozhnyuk2018modelling, hager2014cfd} is used in the current work. The numerical procedure to solve these equations mentioned above within the resolved CFD-DEM framework and the numerical correction step to guarantee the divergence-free condition of the velocity field are detailed below:
\begin{itemize}
\item First, an intermediate velocity field $\hat{\mathbf{U}}$ is solved from the Navier-Stokes equations (Eqs.~\ref{ConEqn-ReCFDEM} and \ref{MomEqn-ReCFDEM}) over the whole computational domain. In this step, the presence of solid particles in the CFD domain is not considered.
\item Second, the intermediate velocity field $\hat{\mathbf{U}}$ in the CFD cells covered by a solid particle is overwritten by imposing the particle velocity calculated from the DEM side, explicitly. This leads to a new velocity field $\tilde{\mathbf{U}}$.
\item In general, the new velocity field $\tilde{\mathbf{U}}$ is not divergence-free. Thus, a numerical correction step is further needed to correct this velocity. A Poisson equation given by 
\begin{equation} \label{velPotEqn-Re}
\nabla^2 \phi_r = \nabla \cdot \tilde{\mathbf{U}} - \epsilon_f \dot{m} (\frac {1}{\rho_g}-\frac {1}{\rho_l})
\end{equation} 
is solved for the velocity potential field $\phi_r$. Another new velocity from numerical correction is defined as $\bar{\mathbf{U}}$ given by
\begin{equation}
\bar{\mathbf{U}} = \tilde{\mathbf{U}} - \nabla \phi_r.
\end{equation}
It can be proven that the new velocity field $\bar{\mathbf{U}}$ is divergence-free:
\begin{equation} \label{velProof}
\begin{split}
\nabla \cdot \bar{\mathbf{U}} = \nabla \cdot \left( \tilde{\mathbf{U}} - \nabla \phi_r \right) & = \nabla \cdot \tilde{\mathbf{U}} - \underbrace{\nabla \cdot \nabla \phi_r}_{=\nabla^2 \phi_r} \\
& = \nabla \cdot \tilde{\mathbf{U}} - \nabla \cdot \tilde{\mathbf{U}} + \epsilon_f \dot{m} (\frac {1}{\rho_g}-\frac {1}{\rho_l}) \\
& = \epsilon_f \dot{m} (\frac {1}{\rho_g}-\frac {1}{\rho_l}).
\end{split}
\end{equation}
Here the last term $\epsilon_f \dot{m} (\frac {1}{\rho_g}-\frac {1}{\rho_l})$ in Eqs.~\ref{velPotEqn-Re} and \ref{velProof} results from the phase change of the liquid phase which is equal to zero when there is no phase change.
\item The velocity potential field $\phi_r$ is also used to correct the pressure field by
\begin{equation}
p = \hat{p} + \rho \frac{\phi_r}{\Delta t},
\end{equation}
where $\hat{p}$ is the pressure field solved from the Navier-Stokes equations (Eqs.~\ref{ConEqn-ReCFDEM} and \ref{MomEqn-ReCFDEM}), $\rho$ is the density field and $\Delta t$ the time step.
\end{itemize}

\subsection{Coupling algorithm}
The open-source framework CFDEMcoupling-PUBLIC \cite{githubDCS} is extended to implement the variable-density-based multiphase framework coupling CFD to DEM. A new coupling solver named cfdemSolverVoFIB based on the standard solver cfdemSolverIB of the CFDEMcoupling-PUBLIC library is developed in this paper. The cfdemSolverIB solver is not capable of modelling variable-density incompressible flow with surface tension and evaporation. Accordingly, the extended solver cfdemSolverVoFIB is developed in this work. The new solver is capable of realizing the following functionality:
\begin{itemize}
\item Get particle data, e.g. particle coordinates, velocity and particle radius from DEM.
\item Identify CFD cells covered by solid particles and calculate the void fraction $\epsilon_f$ of each CFD cell.
\item Correct the velocity field when solid particles are present in the liquid phase.
\item Calculate the fluid-solid interaction force and capillary force.
\item Give essential data to DEM, e.g. buoyancy, capillary force, fluid-solid interaction force, etc.
\item Repeat these steps mentioned above until the simulation ends.
\end{itemize}

\begin{figure}[h]
 \begin{center}
 	\includegraphics[width=1.0\textwidth]{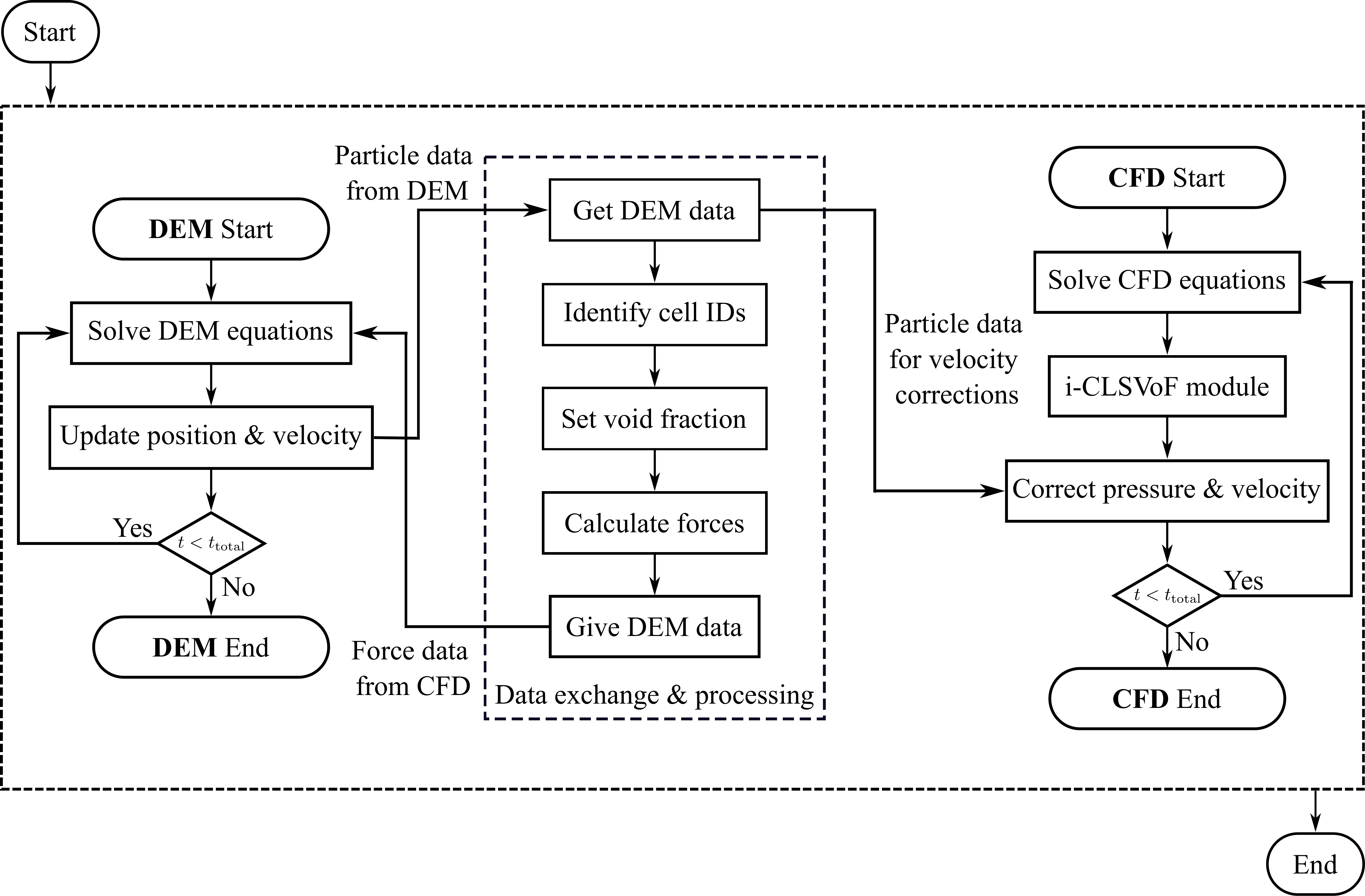}
  \end{center}
  \caption{The coupling algorithm for the resolved CFD-DEM approach.}
  \label{resolvedCFD-DEM_algorithm}
\end{figure}

A detailed coupling algorithm between DEM and CFD for the resolved CFD-DEM approach is shown in Figure~\ref{resolvedCFD-DEM_algorithm}. The solver cfdemSolverVoFIB consists of three modules, namely, the CFD module, the DEM module, and the data exchange and processing module as shown in Figure~\ref{resolvedCFD-DEM_algorithm}.
OpenFOAM and LIGGGHTS are employed to conduct the CFD and DEM simulations, respectively. In the resolved CFD-DEM simulations, DEM and CFD conduct their simulations, separately, once the CFD-DEM simulation starts. In DEM, the governing equations (Eqs.~\ref{DEMGEqn1-ReCFDEM} and \ref{DEMGEqn2-ReCFDEM}) are solved to update the velocity, position and other information for solid particles. These information is transferred to the data exchange and processing module for further calculations as detailed below. In CFD, the governing equations (Eqs.~\ref{ConEqn-ReCFDEM} and \ref{MomEqn-ReCFDEM}) are first solved to update an intermediate velocity field, ignoring solid particles present in the liquid phase. The i-CLSVoF module detailed in our previous work \cite{xia2022improved} is used to capture the moving free surface and calculate the surface tension force.

The data exchange between DEM and CFD is crucial to realize the resolved CFD-DEM coupling. The essential data, e.g. particle positions and velocities calculated on the DEM side, are transferred to the data exchange and processing module. Then, the CFD cells covered by solid particles can be identified, and the void fraction for these CFD cells can be obtained. Furthermore, the particle-fluid interaction forces can be calculated in the data processing module and then transferred to the DEM side to update particle data in the next cycle. Particle velocity data from DEM is transferred to the CFD side, and the numerical correction step is then used to correct the velocity field to satisfy the divergence-free condition as discussed in Section~\ref{NumCorrections}.

The time-step size for stable DEM simulations is given by the Rayleigh time-step given by
\begin{equation} \label{T_R}
{\Delta t}^{\text{DEM}} = f_s \frac{\pi \bar R \sqrt{\frac{2 \rho (1+\nu)}{Y}}}{0.1631 \nu+0.8766},
\end{equation}
where $\bar R$ is the average particle radius, $\rho$ the particle density, $Y$ the Young's modulus and $\nu$ the Poisson's ratio \cite{li2005comparison}. Additionally, $f_s$ is a safety factor for which a value ranging from $0.1$ to $0.3$ is recommended.
The maximum time step for guaranteeing a stable CFD simulation is given by two constrains. The first constraint is 
\begin{equation} \label{deltaT_sigma}
\Delta_{t{\sigma}} < \sqrt \frac{\rho_{\text{avg}} {\Delta x}^3}{2 \pi \sigma},
\end{equation}
where $\rho_{\text{avg}}$ is the average density of the phases. It is proposed for the explicit treatment of the surface tension force term \cite{brackbill1992continuum}. Another more comprehensive time step constraint is given by 
\begin{equation} \label{deltaT_tc}
\Delta_{tc} < \frac{1}{2} \left(C_2 \tau_{\mu} + \sqrt{(C_2 \tau_{\mu})^2+4C_1 {\tau_{\rho}}^2} \right),
\end{equation}
which involves the density and the viscosity of the multiphase system. $\tau_{\mu} $ and $\tau_{\rho}$ are given as $\mu_{\text{avg}} \Delta x /\sigma$ and $\sqrt{\rho_{\text{avg}} \Delta x^3 / \sigma}$, respectively, with $\mu_{\text{avg}}$ being the average dynamic viscosity of the liquid and gas phases \cite{galusinski2008stability}. Accordingly, the maximum time step size for stable CFD simulations is given as
\begin{equation} \label{deltaT_i-CLSVoF}
{\Delta t}^{\text{CFD}} < \text{min}(\Delta_{t\sigma}, \Delta_{tc})C_{\Delta t}
\end{equation}
with $C_{\Delta t}$ being the stabilization factor where a  range of $C_{\Delta t}$ between $0.3$ and $0.7$ is recommended for more stable calculation.

The minimal coupling interval for data exchange between DEM and CFD is defined by
\begin{equation}
i_c = \frac{{\Delta t}^{\text{CFD}}}{{\Delta t}^{\text{DEM}}},
\end{equation}
which must be an integer. Increasing the coupling interval requires less computational cost; however, the coupled simulations may then not be accurate enough as the latest data are not exchanged between DEM and CFD in time. The particle data from DEM is used to calculate the void fraction, fluid-structure interaction force, etc. These interaction forces are given back to DEM, and thus the interaction forces acting on the solid phase by the fluid phase can be obtained. DEM and CFD go to the next loop once one data exchange is completed, and the whole simulation ends until the prescribed total simulation time is reached.

\section{Results and discussion}
\label{Results}
\subsection{Numerical validation}
In this section, two benchmark cases are used to validate the resolved CFD-DEM solver cfdemSolverVoFIB developed in this work. The first case compares the drag coefficient calculated with the resolved CFD-DEM approach against a formula. The other case is to compare the settling velocities and particle position against the corresponding experimental results when a spherical particle settles in liquids.
\subsubsection{Validation of calculations of the drag coefficient}
\begin{figure}[h]
	\begin{center}
    \includegraphics[width=0.2\textwidth]{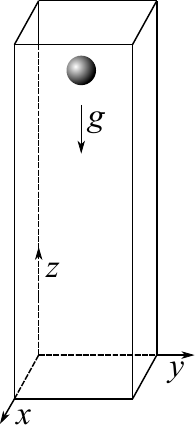}
  \end{center}
  \caption{The schematic diagram of the numerical set-up for calculations of the drag coefficient.}
  \label{diagram_singleSphereSettling}
\end{figure}

The schematic diagram for calculating the drag coefficient is shown in Figure~\ref{diagram_singleSphereSettling}. A sphere falls down under the influence of gravity. 
The sphere is fully immersed in the liquid of a container, and the essential parameters for the numerical simulations are listed in Table~\ref{dragCoeffTable}.
\begin{table}[h]
\centering
\caption{Parameters for numerical simulations used to validate calculations of the drag coefficient.}
\label{dragCoeffTable}
\begin{tabular}{lc}
\hline
Parameter  	& Value [units]	\\ \hline
Domain size (length, width, height)  & $(20, 20, 60)$ [\si{mm}]  \\
Particle diameter   & $2$ [\si{mm}]   \\
Particle density   & $3000$ [\si{kg/m^3}]  \\
Liquid density    & $1000$ [\si{kg/m^3}]  \\
Initial particle position $(x,y,z)$  & $(10, 10, 50.5)$ [\si{mm}] \\ \hline
\end{tabular}
\end{table}
The dynamic viscosity of the liquid in the container significantly influences the sphere's motion, and a wide range of Reynolds numbers can be achieved by varying the liquid dynamic viscosity. Seven numerical benchmark cases with different dynamic viscosities (refer to Table~\ref{dynaVisTable}) are conducted in this section.
\begin{table}[h]
\centering
\footnotesize
\caption{Dynamic viscosities for these seven different cases.}
\label{dynaVisTable}
\bgroup
\def\arraystretch{1.25}
\begin{tabular}{cccccccc}
\hline
Case No. & 1 & 2 & 3 & 4 & 5 & 6 & 7 \\ \hline
$\mu$ [\si{Pa \ s}] & $2.5$ & $5 \times 10^{-1}$ & $1 \times 10^{-1}$ & $5 \times 10^{-2}$ & $1 \times 10^{-2}$ & $5 \times 10^{-3}$ & $3 \times 10^{-3}$ \\ \hline
\end{tabular}
\egroup
\end{table}

The motion of a single sphere inside the container is governed by Newton's second law of motion given by
\begin{equation} \label{govEqn_sphere}
m \frac{d\mathbf{U}}{dt}= \mathbf{F}_g - \mathbf{F}_b - \mathbf{F}_d,
\end{equation}
where $\mathbf{F}_g$, $\mathbf{F}_b$ and $\mathbf{F}_d$ are the gravitational force, buoyancy and drag force acting on the sphere, respectively \cite{norouzi2016coupled}. Substituting expressions of these force terms into Eqn.~\ref{govEqn_sphere}, leads to
\begin{equation}
\frac{\pi \rho_p D_p^3}{6} \frac{d\mathbf{U}}{dt}=\frac{\pi (\rho_p - \rho_f)D_p^3 \mathbf{g}}{6} - \frac{1}{8}C_d \pi D_p^2 \rho_f \mathbf{U}^2,
\end{equation}
where $D_p$ is the particle diameter, and $\rho_p$ and $\rho_f$ are the density for the solid and liquid phases, respectively.

Thus, the rate of change of particle velocity $\mathbf{U}$ with respect to time is given by
\begin{equation} \label{dUdt}
\frac{d\mathbf{U}}{dt}=-\frac{3\rho_f C_d}{4\rho_p D_p}\mathbf{U}^2+ \frac{\rho_p-\rho_f}{\rho_p}\mathbf{g}.
\end{equation}

Typically, the particle velocity increases gradually and reaches a steady velocity, known as the terminal velocity, when a particle settles in a fluid. Thus, the drag coefficient $C_d$ in Eqn.~\ref{dUdt} can be calculated by
\begin{equation} \label{Cd_num}
C_d = \frac{4}{3} \frac{\rho_p-\rho_f}{\rho_f} \frac{|\mathbf{g}| D_p}{|\mathbf{U}_t|^2},
\end{equation}
where $|\mathbf{U}_t|$ is the magnitude of the terminal velocity. Furthermore, the particle Reynolds number is given by
\begin{equation}
\text{Re} = \frac{|\mathbf{U}_t|D_p}{\nu},
\end{equation}
where $\nu$ is the kinematic viscosity of the liquid phase.

Brown et al. corrected the drag coefficient by comparing the corrected formula
\begin{equation} \label{BrownAndLawler}
C_d=\frac{24}{\text{Re}}(1.0 + 0.15 \text{Re}^{0.681})+\frac{0.407}{1+\frac{8710}{\text{Re}}}
\end{equation}
against extensive experimental data \cite{brown2003sphere}. This corrected drag coefficient (denoted as $C_d-\text{Exp}$) is used to validate the drag coefficient calculations in this section where the drag coefficient calculated from the numerical simulations (with Eqn.~\ref{Cd_num}) is denoted as $C_d-\text{Num}$. 

\begin{figure}[h]
	\begin{center}
    \includegraphics[width=0.65\textwidth]{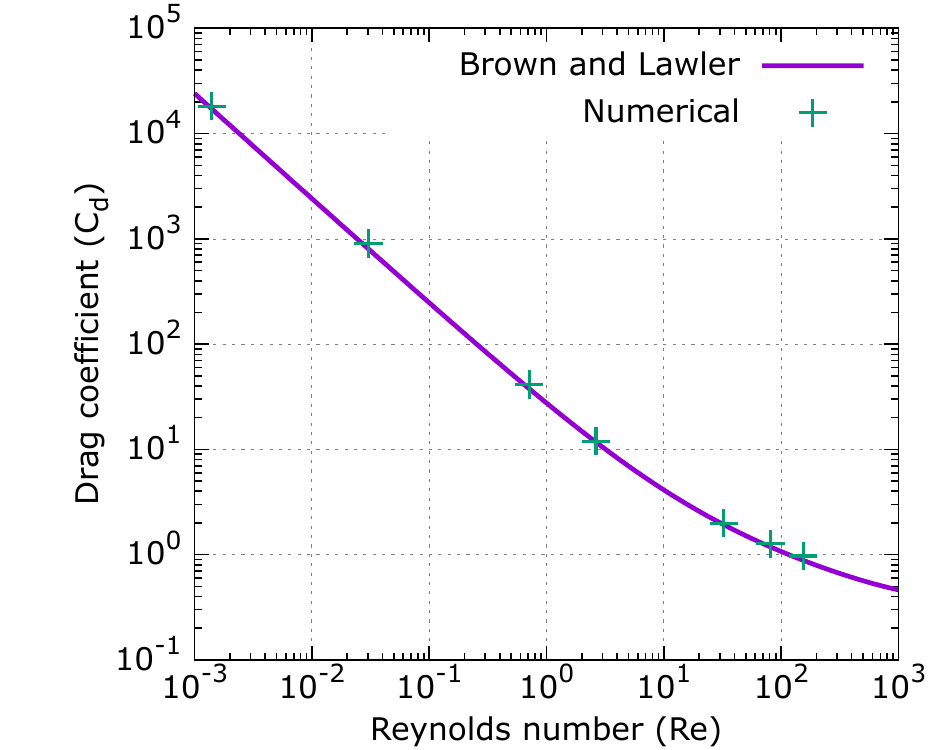}
  \end{center}
  \caption{Validations for calculations of the drag coefficient $C_d$.}
  \label{dragCoefficientValidations}
\end{figure}

The comparison between the numerical drag coefficient and the corrected drag coefficient given by the formula Eqn.~\ref{BrownAndLawler} is shown in Figure~\ref{dragCoefficientValidations}.
Good agreement is obtained for a wide range of Reynolds numbers ranging from  $10^{-3}$ to $10^{3}$. The relative errors of calculating the drag coefficient are listed in Table~\ref{Results4Cases} for the quantitative comparison.
\begin{table}[h]
\centering
\footnotesize
\caption{Terminal velocity and relative error for these seven validation cases.}
\label{Results4Cases}
\bgroup
\def\arraystretch{1.25}
\begin{tabular}{ccccc}
\hline
$\text{Re}$ [-]  & Terminal velocity	[\si{m/s}] & $C_d-\text{Num}$ [-]  & $C_d-\text{Exp}$ [-] & Relative error [-] \\ \hline
0.0014 & 0.0017 & 18103.8062 & 17172.1450 & 0.0543 \\
0.0304 & 0.0076 & 905.8172 & 800.4452 & 0.1316 \\
0.7120 & 0.0356 & 41.2827 & 37.7199 & 0.0945 \\
2.6400 & 0.0660 & 12.0110 & 11.7323 & 0.0238 \\
32.3600 & 0.1618 & 1.9985 & 1.9306 & 0.0352 \\
81.2000 & 0.2030 & 1.2696 & 1.1848 & 0.0716 \\
154.4667 & 0.2317 & 0.9746 & 0.8837 & 0.1029 \\
\hline
\end{tabular}
\egroup
\end{table}
Calculations of the drag coefficient are more accurate for the Reynolds number between $0.1$ and $100$. However, the accuracy of the model needed to be improved for high Reynolds numbers, namely when $\text{Re}$ is larger than $100$. This relatively large discrepancy for simulations with high Reynolds numbers was reported in the literature as well \cite{shen2022resolved, schnorr2022resolved}.

\subsubsection{Single particle settling in a container}
The experimental study and corresponding numerical simulations of single particle settling in viscous liquids were conducted by Ten Cate et al. \cite{ten2002particle}. The experimental set-up for the single particle settling is a spherical bearing ball with a diameter of $15$ \si{mm}, and a density of $1120$ \si{kg/m^3} which settles in a container (length $\times$ width $\times$ height= $100 \times 100 \times 160$ \si{mm}) under the influence of gravity ($g = 9.81$ \si{m/s^2} in the vertical direction). The sphere is fully immersed in the liquid before it starts to fall, and the initial separation distance between the sphere center and the bottom wall of the container is $120$ \si{mm}. The sphere experiences acceleration at the beginning and then decelerates when it approaches the bottom wall. This scenario is suitable for validating the numerical implementations for computing the fluid-solid interaction forces and the trajectory of the spherical particle. 

In this section, four cases with different liquid densities and dynamic viscosities are considered to validate the numerical model developed in this paper. These essential parameters are detailed in Table~\ref{paras4Exps}. 
\begin{table}[h]
\centering
\footnotesize
\caption{Parameters for the single particle settling simulations (data adopted from \cite{ten2002particle}).}
\label{paras4Exps}
\bgroup
\def\arraystretch{1.25}
\begin{tabular}{cccc}
\hline
Case No.  & $\text{Re}$ [-] & Liquid density [\si{kg/m^3}] & Dynamic viscosity [\si{Pa \ s}] \\ \hline
1 & 1.5 & 970 & 0.373 \\
2 & 4.1 & 965 & 0.212 \\ 
3 & 11.6 & 962 & 0.113 \\
4 & 31.9 & 960 & 0.058 \\ \hline
\end{tabular}
\egroup
\end{table}
The number of cells of the base mesh resolution is $40 \times 40 \times 64$. Adaptive mesh refinement is used to guarantee fine mesh resolution around the sphere and to allow a relatively coarse mesh elsewhere to reduce computational cost while guaranteeing reasonable numerical accuracy. The no-slip boundary condition is applied to the boundary of the container. The time-step size for both DEM and CFD is $1.0 \ \times 10^{-5}$, and a coupling interval of one is used to exchange data between DEM and CFD.

\begin{figure}[h]
	\begin{center}
    \includegraphics[width=0.35\textwidth]{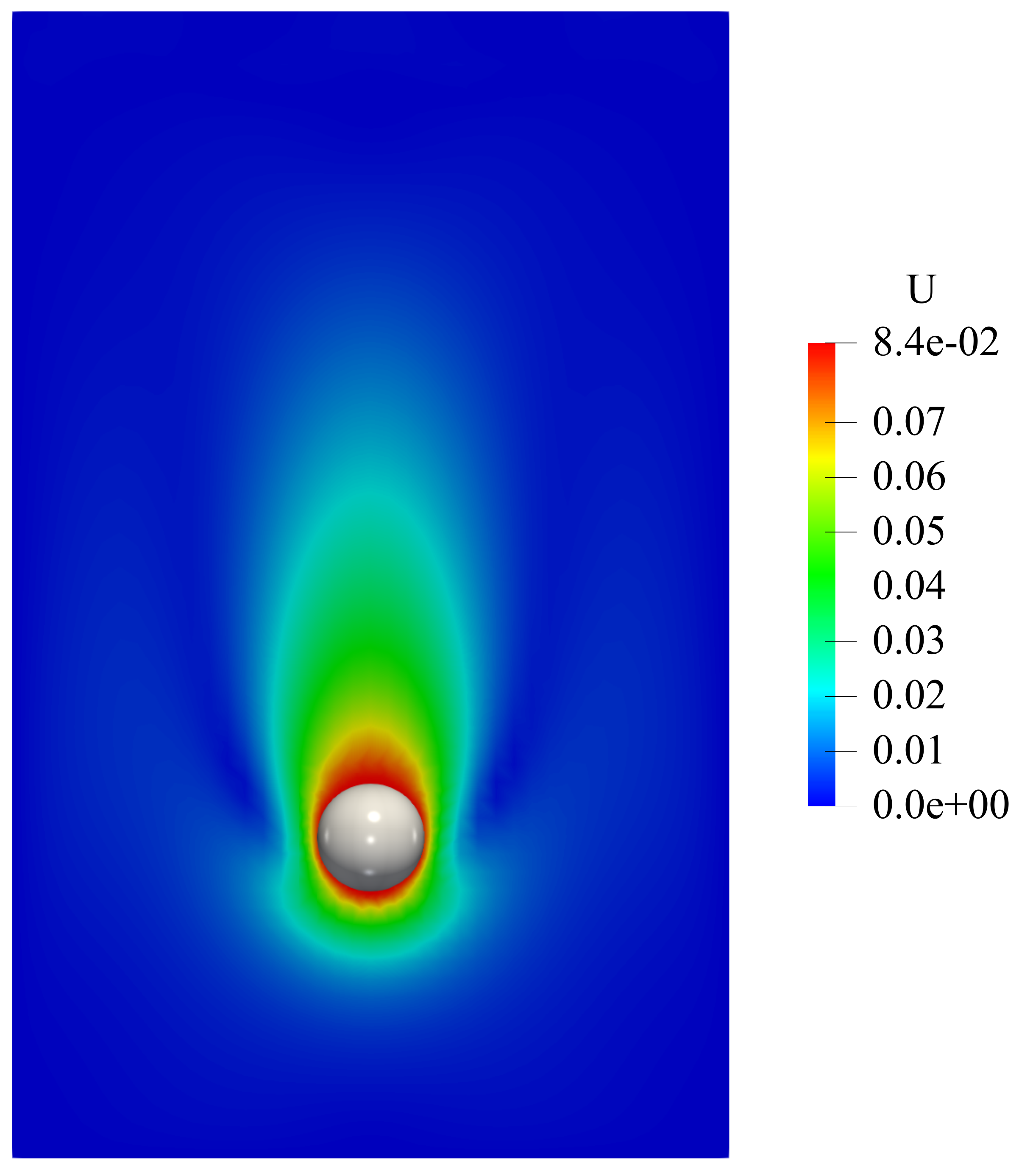}
  \end{center}
  \caption{The liquid velocity field of single particle settling simulation ($\text{Re} = 11.6$).}
  \label{singleSphereVelField}
\end{figure}

\begin{figure}[h]
\centering
  \begin{subfigure}[h]{0.495\textwidth}
    \includegraphics[width=\textwidth]{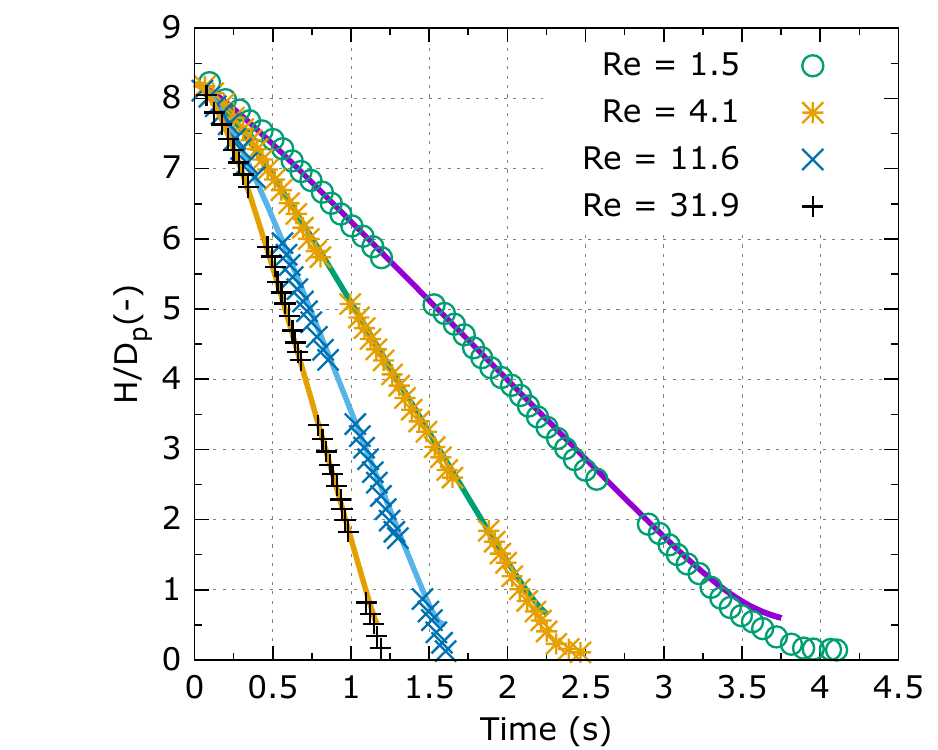}
    \caption{}
    \label{sphereDimlessH}
  \end{subfigure}
  \hfill
  \begin{subfigure}[h]{0.495\textwidth}
    \includegraphics[width=\textwidth]{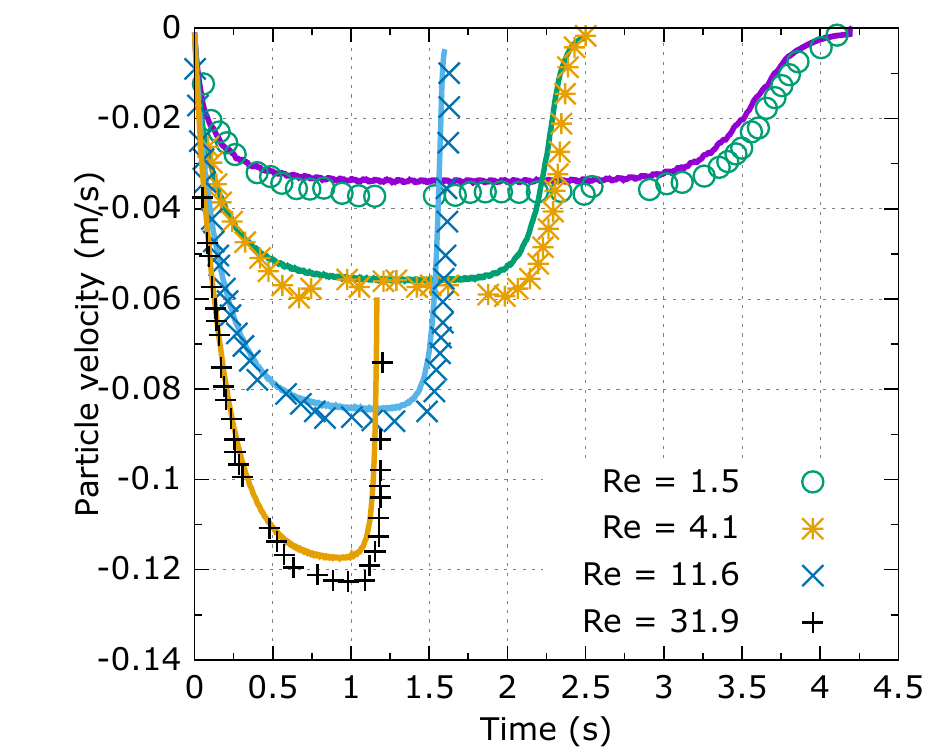}
    \caption{}
    \label{sphereVel}
  \end{subfigure}
  \caption{Validations of the single sphere settling in liquids: (a) dimensionless sphere height, (b) sphere settling velocity.}
\label{singleSphereSettlingValidation}
\end{figure}

The liquid velocity field for $\text{Re}=11.6$ is shown in Figure~\ref{singleSphereVelField}. 
An elongated wake can be seen from the simulation. The dimensionless gap height $H/D_p$ between the sphere and the bottom wall and the magnitude of the particle settling velocity in the vertical direction are recorded and compared against the corresponding experimental results. This comparison is shown in Figure~\ref{singleSphereSettlingValidation}. 
Figure~\ref{sphereDimlessH} and Figure~\ref{sphereVel} are the dimensionless gap height and particle settling velocity, respectively. The solid curves represent numerical results, while the points represent experimental results adopted from the literature \cite{ten2002particle}. As shown in Figure~\ref{sphereVel}, the spherical particle first undergoes an acceleration phase and then decelerates due to squeezing liquid between the sphere and the bottom wall when the particle approaches the bottom wall. A good agreement can be found for these four different cases. Some minor discrepancies can be seen from Figure~\ref{sphereDimlessH}, especially when $\text{Re} = 1.5$. The reason is that the lubrication force is not negligible when the gap between the particle and the bottom wall is small. Incorporating the lubrication force can be subject to future work which is not included in the current work. Overall, these simulations demonstrate that the numerical implementation for the resolved CFD-DEM model is correct, and that the model is accurate enough to capture the complex fluid-solid interaction and to predict the trajectory of solid particles.

\subsection{Application}
The improved resolved CFD-DEM model developed in this work can be used to model capillary-force-induced or evaporation-induced transport and agglomeration of particles. In this section, two numerical benchmark cases are presented to demonstrate the performance of the resolved CFD-DEM model with the capillary interactions developed in this paper.
\subsubsection{Two particles moving along a free surface with evaporation}
\label{liquidBridge}
\begin{figure}[h]
  \begin{center}
    \includegraphics[width=0.65\textwidth]{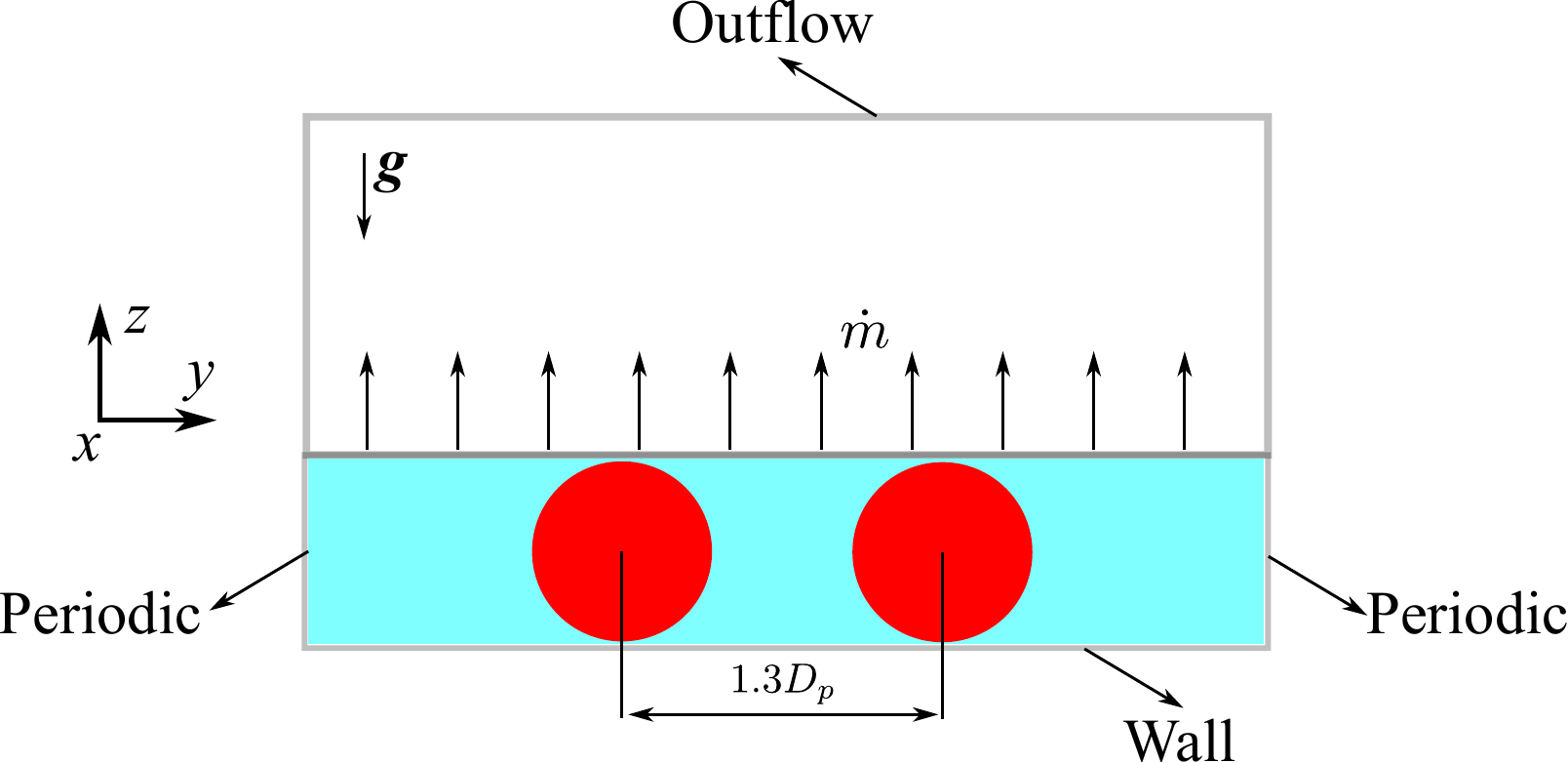}
  \end{center}
  \caption{2D schematic diagram of the numerical set-up for two particles moving along a free surface with evaporation.}
\label{2ParticleEvap}
\end{figure}

The first simulation is to model the evaporation-induced deformation of the meniscus between two spherical particles sitting on a substrate. The numerical set-up for the 3D simulation is shown in Figure~\ref{2ParticleEvap}, namely, two spherical particles resting on a wettable substrate. The initial distance between the centers of the two particle is $1.3 D_p$ (particle diameter).
Periodic boundary conditions are applied in $x$ and $y$ directions. The initial liquid height is higher than the particle diameter. The outflow boundary condition is applied at the top to let vapour leave the domain freely. The constant contact angle and slip boundary conditions are applied on the bottom wall. The specified contact angle between the spherical particle and the liquid surface is $30^{\circ}$, and a constant contact angle specified at the bottom wall is $0^{\circ}$. The constant mass flux evaporation model discussed in our previous work \cite{xia2022improved} is adopted to model the evaporation of the liquid phase into the gas phase with a constant evaporation rate in this section. To demonstrate the performance of the capillary force model extended in this paper, only the gravitational force, buoyancy and capillary force are effective, while the particle-liquid interaction force given by Eqn.~\ref{F_DragCal2} is not considered in the numerical simulation. The particle-liquid interaction force also influences the movements of the two particles during the evaporation process. If this force is active, it is impossible to determine the lateral movement of the two particles due to the capillary force only. The parameters for this simulation are listed in Table~\ref{Paras4TwoPEvap}.
\begin{table}[h]
\centering
\caption{Parameters for two particles moving along a free surface with evaporation.}
\label{Paras4TwoPEvap}
\begin{tabular}{lc}
\hline
Parameter [Units]               & Value \\ \hline
Liquid density [\si{kg/m^3}]         & 10  \\
Gas density [\si{kg/m^3}]             & 1     \\
Particle density [\si{kg/m^3}]       & 25     \\
Particle diameter [\si{m}]        & $1.0 \times 10^{-6}$     \\
Liquid viscosity [\si{Pa \ s}]      & $1.0 \times 10^{-3}$  \\
Gas viscosity [\si{Pa \ s}]           & $1.0 \times 10^{-5}$  \\
Surface tension [\si{N/m}]        & 0.072   \\
CFD time step [\si{s}]        & $1.0 \times 10^{-9}$  \\
DEM time step [\si{s}]          & $1.0 \times 10^{-9}$  \\
Coupling interval $[-]$      & 1   \\
Restitution coefficient $[-]$ & 0.5   \\
Friction coefficient $[-]$   & 0.3   \\
Contact angle (particle-interface) $[-]$   & $30^{\circ}$, $45^{\circ}$, $60^{\circ}$   \\ Contact angle (particle-wall) $[-]$   & $0^{\circ}$ \\ \hline 
\end{tabular}
\end{table}

\begin{figure}[h]
\centering
  \begin{subfigure}[h]{0.32\textwidth}
    \includegraphics[width=\textwidth]{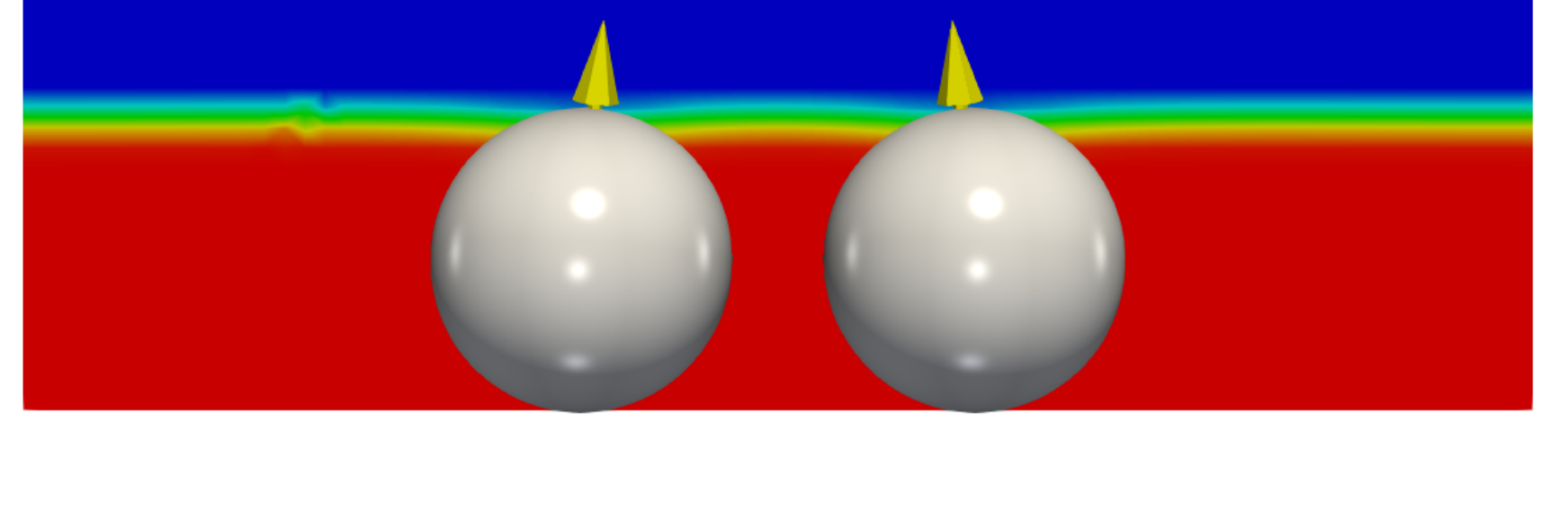}
    \caption{}
    \label{30Deg_vector01}
  \end{subfigure}
  \hfill
  \begin{subfigure}[h]{0.32\textwidth}
    \includegraphics[width=\textwidth]{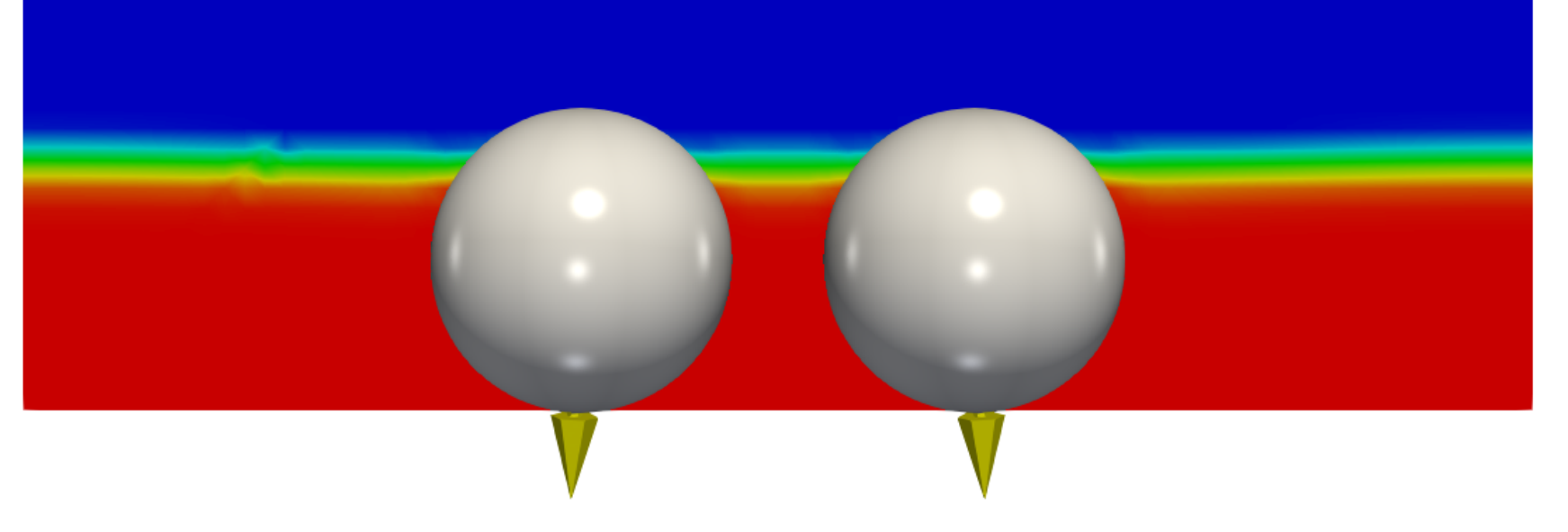}
    \caption{}
    \label{30Deg_vector02}
  \end{subfigure}
  \hfill
    \begin{subfigure}[h]{0.32\textwidth}
    \includegraphics[width=\textwidth]{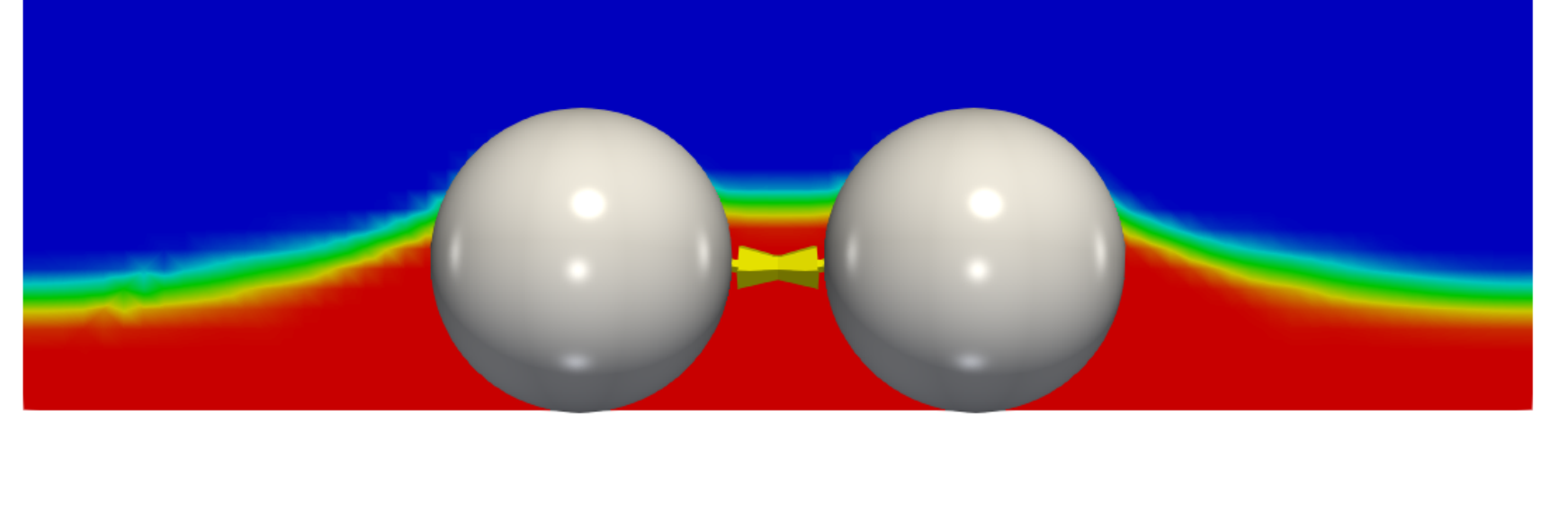}
    \caption{}
    \label{30Deg_vector03}
  \end{subfigure}
  \caption{Vectors of the particle velocity are represented by the yellow arrows during the evaporation process: (a) pointing upwards, (b) pointing downwards, (c) pointing towards each other.}
\label{30Deg_vector}
\end{figure}

Two particles gradually protrude from the liquid surface after evaporating some liquid from the liquid surface, as shown in Figure~\ref{30Deg_vector}. 
The vectors of the particle velocity point upwards during the early stage of evaporation (see Figure~\ref{30Deg_vector01}). This is due to the upward capillary force acting on the two particles when they protrude from the liquid surface. Then, the vectors of the particle velocity point downwards (see Figure~\ref{30Deg_vector02}) after evaporating more liquid. A concave meniscus between the two particles gradually forms, which leads to attractive interactions between them, as shown in Figure~\ref{30Deg_vector03}. This attractive capillary force acting on the two particles makes them moving toward each other.

\begin{figure}[h]
  \begin{center}
    \includegraphics[width=1.0\textwidth]{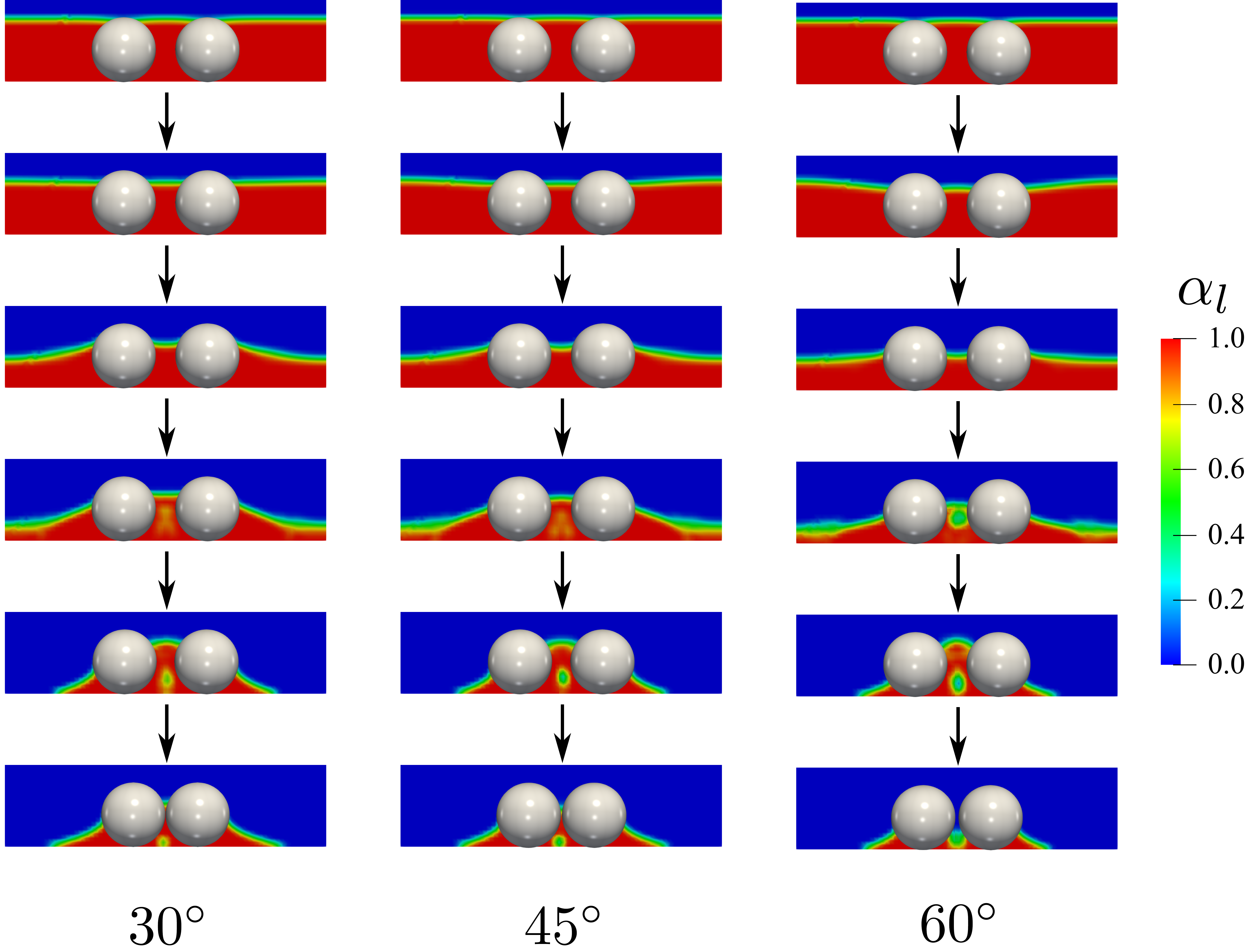}
  \end{center}
  \caption{Numerical simulations of two particles moving along the free surface for three different specified contact angles.}
\label{2Particles_liquidBridge}
\end{figure}

Three simulations with different contact angles, namely, $30^{\circ}$, $45^{\circ}$ and $60^{\circ}$ are presented in Figure~\ref{2Particles_liquidBridge}. The parameters for the solid and liquid phases can be found in Table~\ref{Paras4TwoPEvap}.
For the very early stage of evaporation, the free surface gradually decreases due to the mass loss and then contacts with the top of the spheres. The virtual free surface inside the solid particles is concave upward, as shown in the top row of Figure~\ref{2Particles_liquidBridge}. More mass loss can be found around the two particles when the contact angle increases from $30^{\circ}$ to $60^{\circ}$, as shown in the second row of Figure~\ref{2Particles_liquidBridge}. A concave meniscus and a liquid bridge gradually form around every two particles for the cases with contact angles $\theta = 30^{\circ}$ and $\theta = 45^{\circ}$, while a flat meniscus is found for the case with a contact angle of $60^{\circ}$, as shown in the third row of Figure~\ref{2Particles_liquidBridge}. These concave meniscuses lead to attractive capillary forces and force each pair of particles to come closer to each other. As the simulation continues, more liquid evaporates around two sides of the computational domain, and the shape of the meniscus changes from concave to convex, as demonstrated by the fourth and fifth rows of Figure~\ref{2Particles_liquidBridge}. In all three cases, particles gradually move towards each other during the evaporation process. The liquid phase evaporates faster when the contact angle increases from $30^{\circ}$ to $60^{\circ}$. The same conclusion is also shown in Fig. 4 presented in the literature \cite{mino2022numerical}. 

\subsubsection{Particle transport and accumulation in an evaporating droplet with contact line pinning}
In contrast to the aforementioned numerical simulations, the current numerical benchmark case involves many particles inside an evaporating droplet with contact line pinning. In principle, droplet evaporation with contact line pinning results in an internal capillary flow from the droplet center to the edge \cite{deegan1997capillary}. The radial capillary flow carries some suspended particles from the droplet center to its edge and finally leads to an inhomogeneous particle deposition pattern. Accordingly, the internal flow field is of great significance in affecting particle transport and accumulation during the evaporation process. This is in contrast to the first numerical demonstration case as discussed in Section~\ref{liquidBridge}, where capillary force is more dominant, and the internal velocity field does not play a significant role. 

\begin{figure}[h]
  \begin{center}
    \includegraphics[width=0.5\textwidth]{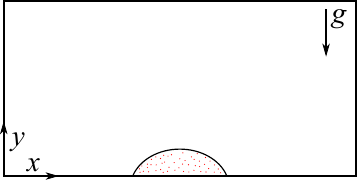}
  \end{center}
  \caption{The schematic diagram of the numerical setup for 2D droplet evaporation with suspended particles.}
  \label{2DCRE_numericalSetup}
\end{figure}

In order to save computational cost and to visualize the evaporation-induced particle transport during the evaporation process, a 2D numerical simulation is adopted in this section, namely, all the particles inside the evaporating droplet can only move along the $x$ and $y$ directions. The numerical set-up is shown in Figure~\ref{2DCRE_numericalSetup}. 
$300$ micro-sized spherical particles with a diameter of $1$ \si{\mu m} are generated randomly inside a spherical cap with an initial contact angle of $45^{\circ}$ and a radius of $50$ \si{\mu m}. The length and height of the computational domain are $300$ \si{\mu m} and $150$ \si{\mu m}, respectively. As the particle size is small, the Van der Waals force model detailed in the Appendix is incorporated to account for the non-contacting attractive force acting on the solid particles during the evaporation process. The parameters used in the simulation are listed in Table~\ref{Paras4CRE}.
\begin{table}[h]
\centering
\caption{Essential parameters for modelling particle transport inside an evaporating droplet.}
\label{Paras4CRE}
\begin{tabular}{lc}
\hline
Parameter [Units]               & Value \\ \hline
Liquid density [\si{kg/m^3}]         & 10  \\
Gas density [\si{kg/m^3}]             & 1     \\
Particle density [\si{kg/m^3}]        & 250     \\
Particle diameter [\si{m}]        & $1.0 \times 10^{-6}$     \\
Liquid viscosity [\si{Pa \ s}]      & $1.0 \times 10^{-3}$  \\
Gas viscosity [\si{Pa \ s}]          & $1.0 \times 10^{-5}$  \\
Surface tension [\si{N/m}]        & 0.072   \\
Surface energy density [\si{J/m^2}]        & $0.86 \times 10^{-3}$   \\
CFD time step [\si{s}]        & $1.0 \times 10^{-12}$  \\
DEM time step [\si{s}]          & $1.0 \times 10^{-12}$  \\
Coupling interval $[-]$      & 1   \\
Restitution coefficient $[-]$ & 0.5   \\
Friction coefficient $[-]$   & 0.3   \\
Contact angle (particle-interface) $[-]$   & $30^{\circ}$   \\ Contact angle (particle-wall) $[-]$   & $0^{\circ}$ \\ \hline 
\end{tabular}
\end{table}

The no-slip boundary condition is applied at the bottom wall to fix the contact line during the evaporation process, and the outflow boundary condition is applied at the top to let vapour leave the domain freely. In order to speed up the numerical simulations and mitigate the influence of un-physical spurious velocities on the internal flow field inside the evaporating droplet, the density of the liquid phase is scaled by $0.01$. The density-scaled approach is also adopted to model droplet evaporation in the literature \cite{ledesma2014lattice, irfan2017front, zhang2021immersed}. A 2D axisymmetrical model was used in the literature \cite{zhang2021immersed}; however, a non-symmetrical numerical configuration, as shown in Figure~\ref{2DCRE_numericalSetup}, is used in this section. The reason is that the initial particle packing for the DEM simulations is not axisymmetric. 

\begin{figure}[h]
\centering
  \begin{subfigure}[h]{0.65\textwidth}
    \includegraphics[width=\textwidth]{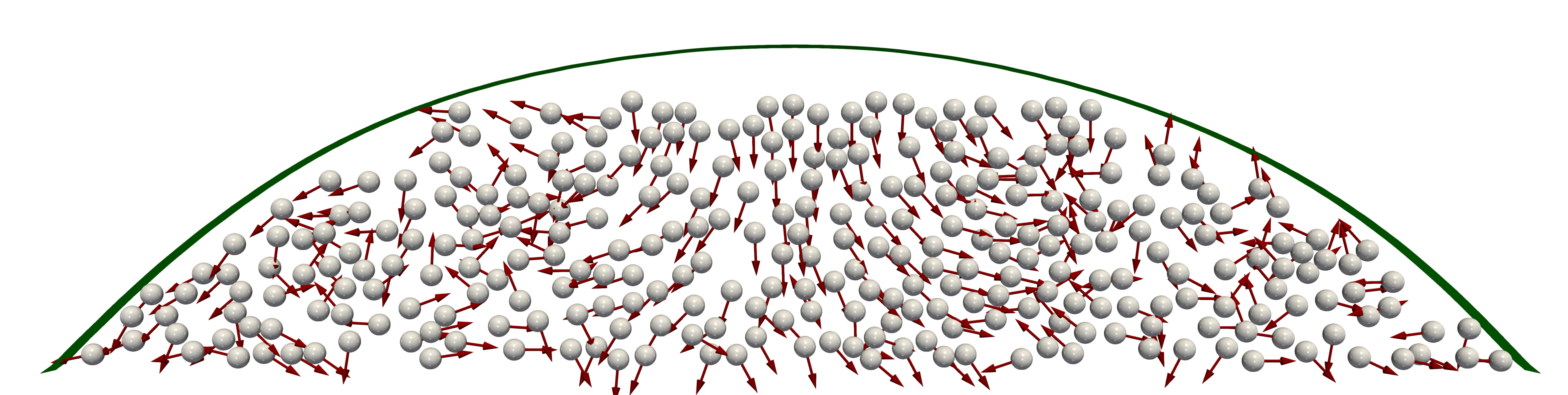}
    \caption{}
    \label{2DCRE_01}
  \end{subfigure}
  \hfill
  \begin{subfigure}[h]{0.65\textwidth}
    \includegraphics[width=\textwidth]{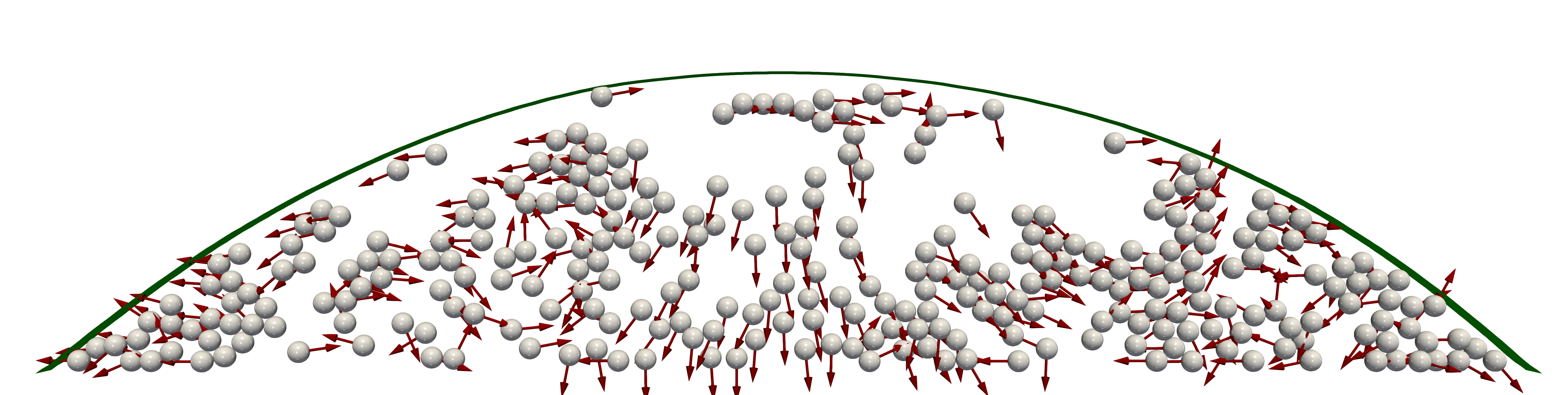}
    \caption{}
    \label{2DCRE_02}
  \end{subfigure}
  \hfill
    \begin{subfigure}[h]{0.65\textwidth}
    \includegraphics[width=\textwidth]{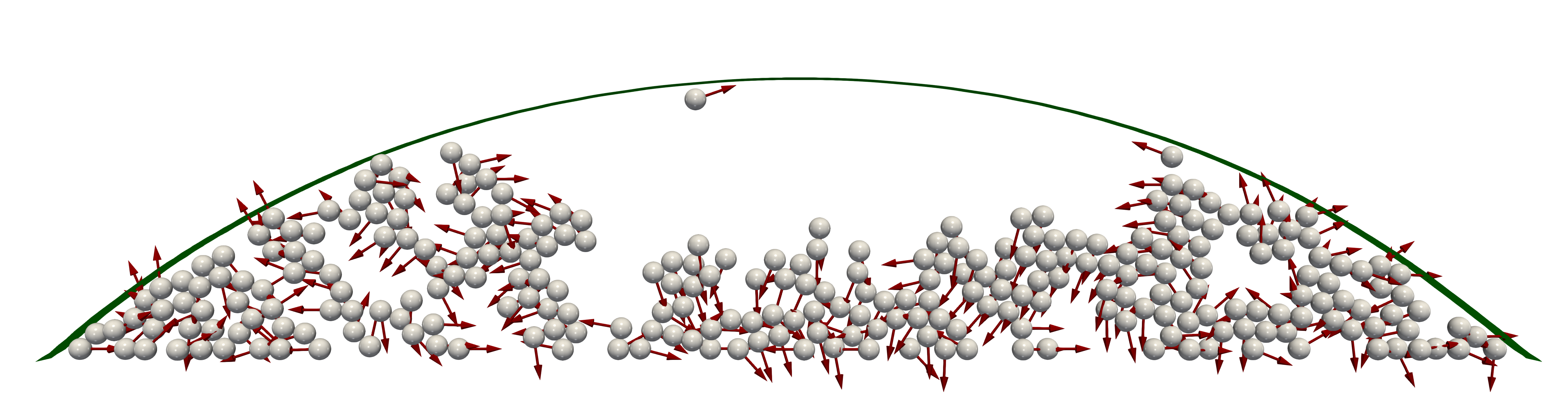}
    \caption{}
    \label{2DCRE_03}
  \end{subfigure}
    \hfill
    \begin{subfigure}[h]{0.65\textwidth}
    \includegraphics[width=\textwidth]{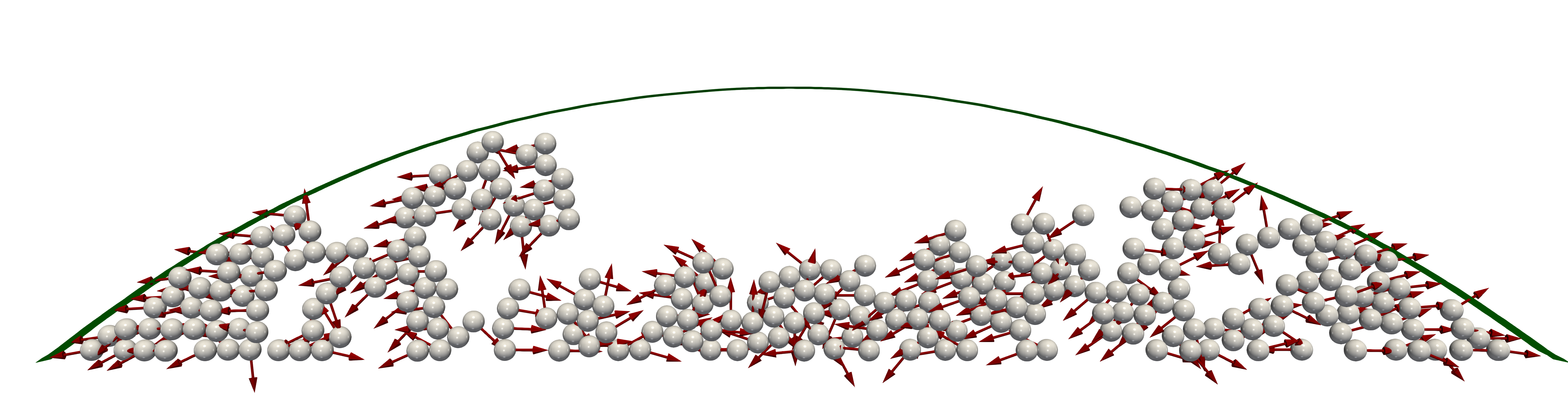}
    \caption{}
    \label{2DCRE_04}
  \end{subfigure}
  \caption{Snapshots of the numerical simulations of the four stages of particle deposition patterns inside an evaporating droplet.}
\label{2DCRE_simus}
\end{figure}

Figure~\ref{2DCRE_simus} shows snapshots of the numerical simulations. The surface in green represents the free surface of the evaporating droplet.
As the evaporation proceeds, the initial contact radius between the sessile droplet and the substrate is constant, and the droplet height decreases gradually. The red arrows indicate the velocity vectors of particles. In the initial stage of the evaporation as shown in Figure~\ref{2DCRE_01}, particles in the middle of the droplet tend to move downwards, while particles around the two corners tend to move towards the triple contact line region due to the radial capillary flow. As shown in Figure~\ref{2DCRE_02}, some particles tend to agglomerate with their neighbouring particles around the free surface. More and more particles are dragged towards the two corners of the evaporating sessile droplet as the evaporation proceeds, as shown in Figure~\ref{2DCRE_03}. Figure~\ref{2DCRE_04} shows agglomerations of particles around the triple contact line region, while fewer particles are deposited in the middle of the droplet. 

\begin{figure}[h]
  \begin{center}
    \includegraphics[width=1.0\textwidth]{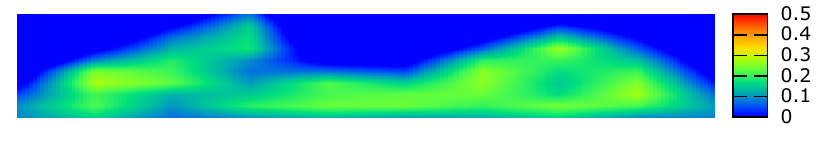}
  \end{center}
  \caption{The local packing fraction for the particle deposition pattern shown in Figure~\ref{2DCRE_04}.}
  \label{packing}
\end{figure}

In this work, a simple yet helpful approach has been developed to calculate the local packing fraction with the open-source Voronoi tessellation code Voro++ (refer to Appendix). This approach is adopted to calculate the local packing structure and packing fraction for the particle assembly in this section. Figure~\ref{packing} shows the local packing fraction for the particle deposition pattern shown in Figure~\ref{2DCRE_04}.
It demonstrates that a higher packing fraction can be found around the two corners where much more particles are agglomerated.

This numerical benchmark case demonstrates that the resolved CFD-DEM model can capture the complex particle-fluid, particle-particle and particle-wall interactions when the liquid phase undergoes phase change from liquid to vapour. In experiments, changing liquid parameters, e.g. the liquid viscosity, was found to alter particle deposition patterns as discussed in the experimental work \cite{mikolajek2018control}. Thus, extensive numerical simulations with this resolved CFD-DEM model can help to understand the effect of certain parameters on the resulting particle deposition patterns in future work.

\section{Conclusions}
\label{Conclusion}
In this paper, the theory and numerical issues of the improved resolved CFD-DEM approach are discussed. A variable-density resolved model is developed, implemented and validated. An improved capillary force model has been implemented into the open-source CFDEMcoupling-PUBLIC library. A corresponding improved resolved CFD-DEM solver cfdemSolverVoFIB has been thus developed. The main contribution and improvements in this work are as follows:
\begin{itemize}
\item A variable-density resolved CFD-DEM model with free-surface capturing has been developed.
\item The evaporation models discussed in our previous work have been incorporated into the cfdemSolverVoFIB solver to model the gas-liquid-solid multiphase system with evaporation of the liquid phase.
\item An improved capillary force model with numerical smoothing has been developed to improve numerical stability issues when modelling capillary interactions for solid particles moving at a free surface.
\end{itemize}
A brief comparison between the standard resolved CFD-DEM solver cfdemSolverIB implemented in the CFDEMcoupling-PUBLIC library and the cfdemSolverVoFIB developed in this work is outlined in Table~\ref{resolvedCFDEM-XIA}.
\begin{table}[h]
\centering
\caption{Comparison between the standard cfdemSolverIB solver and the improved solver cfdemSolverVoFIB.}
\label{resolvedCFDEM-XIA}
\begin{tabular}{lcc}
\hline
Model/Module       & cfdemSolverIB & cfdemSolverVoFIB \\ \hline
Incompressible flow solver              & yes      & yes   \\
Variable density & no      & yes   \\
Free surface capturing      & no      & yes   \\
Surface tension      & no      & yes   \\
Capillary force     & no      & yes   \\ 
Evaporation of the liquid phase    & no      & yes   \\ \hline
\end{tabular}
\end{table}
It turns out that the new solver cfdemSolverVoFIB developed in this paper extends the applications of the standard resolved CFD-DEM solver cfdemSolverIB. 

Two numerical validation cases have been conducted to validate the resolved CFD-DEM solver developed in this work. It is proven that the resolved CFD-DEM solver predicts complex particle-fluid interactions with reasonable numerical accuracy. Two numerical benchmark cases, e.g. two particles moving along a free surface with evaporation and evaporation-induced agglomerations of many particles inside an evaporating droplet, have been presented in this paper. It demonstrates that the performance of the improved resolved CFD-DEM solver is reasonably good in modelling gas-liquid-solid multiphase systems. 

\section * {Acknowledgements}
We sincerely thank the funding from China Scholarship Council (CSC) for the financial support (CSC201808350108), and the Helmholtz Association in Germany. Some simulations were done using the computational source of the BwUniCluster 2.0.

\setcounter{figure}{0} 
\renewcommand\thefigure{A.\arabic{figure}}

\setcounter{equation}{0}
\renewcommand{\theequation}{A.\arabic{equation}}

\setcounter{table}{0}
\renewcommand{\thetable}{A.\arabic{table}}
\section *{Appendix}
\subsection*{The Van der Waals force}
\label{VDWModel}
The Van der Waals force serves as an attractive force between two interacting particles or a particle and a wall. The magnitude of the Van der Waals force between two particles is related to their respective radius $R_i$ and $R_j$, separation distance $h$ and the material property defined by the Hamaker constant $H_a$ \cite{hamaker1937london}. The formula for calculating the Van der Waals force between two particles is defined by
\begin{equation} \label{VdWP2P}
\mathbf{F}_{ij}^{\text{vdw}}=-\frac{H_a}{6}\frac{64R_i^3R_j^3(h+R_i+R_j)}{(h^2+2R_ih+2R_jh)^2(h^2+2R_ih+2R_jh+4R_iR_j)^2},
\end{equation}
where the minus sign means that the force is attractive \cite{yang2000computer}. Additionally, the Van der Waals force between a particle and a wall is given by
\begin{equation} \label{VdWP2W}
\mathbf{F}_{\text{pw}}^{\text{vdw}}=-\frac{H_aR_i}{6h^2},
\end{equation}
where the magnitude of the force depends on particle radius $R_i$, the materials property and the separation distance $h$ between the particle and the wall \cite{abbasfard2016effect}. The crucial material property Hamaker constant $H_a$ is related to the surface energy density $\gamma_s$ and the cutoff distance $h_{\text{min}}$ \cite{gotzinger2003dispersive}. Accordingly, $H_a$ can be calculated by 
\begin{equation} \label{Ha}
H_a = 24 \pi \gamma_s h_{\text{min}}^2.
\end{equation}
Furthermore, the separation distance $h$ appears in the denominator, for which a cutoff distance $h_{\text{min}}$ is defined to avoid numerical singularities when $h$ approaches zero. In this work, $h=h_{\text{min}}$ when $h$ is smaller than $h_{\text{min}}$. 

Corresponding numerical calculations are conducted to validate the numerical accuracy in computational modelling of the attractive force due to the presence of the Van der Waals force, only. 
\begin{table}[h]
\centering
\caption{Parameters for validations of the Van der Waals force model.}
\label{VdWTable}
\begin{tabular}{lc}
\hline
Parameter [Units]  	& Value	\\ \hline
$R_i$ [\si{m}]       & $5 \times 10 ^{-6}$  \\
$R_j$ [\si{m}]       & $5 \times 10 ^{-6}$   \\
$\rho$ [\si{kg/m^3}]    & $2500$  \\
$h_{\text{min}}$ [\si{m}]   & $1.0 \times 10 ^{-8}$  \\
$\gamma_s$ [\si{J/m^2}]  & $0.86 \times 10^{-3}$ \\
$Y$ [\si{kg/(m \cdot s^2)}]         & $1.0 \times 10^{7}$  \\
$\nu$ [-]      & $0.29$  \\ \hline
\end{tabular}
\end{table}
The basic numerical set up is a single particle $i$ approaching another particle $j$ or a fixed wall with an initial separation distance $h_0 \ (h_0 > 0)$ while the Van der Waals forces are recorded for the two cases, respectively. The parameters used in the numerical simulations are outlined in Table~\ref{VdWTable}. The Van der Waals forces collected from the numerical simulations are compared to the analytical solution given by Eqs.~\ref{VdWP2P} and \ref{VdWP2W}, respectively. 
\begin{figure}[h]
\centering
  \begin{subfigure}[h]{0.495\textwidth}
    \includegraphics[width=\textwidth]{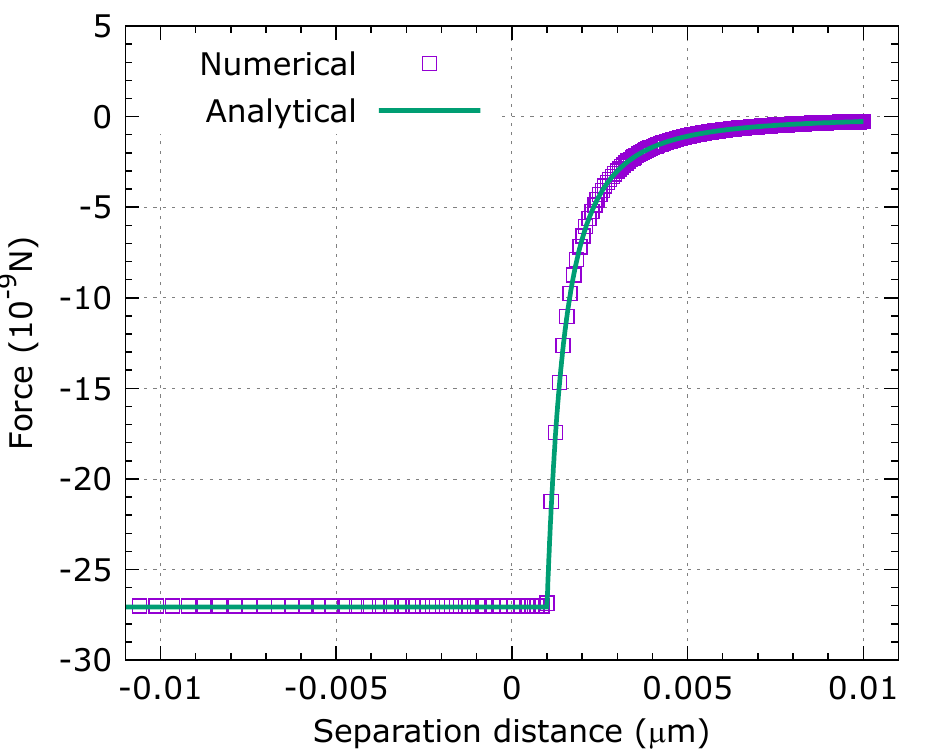}
    \caption{}
    \label{vdw_validationP2P}
  \end{subfigure}
  \hfill
  \begin{subfigure}[h]{0.495\textwidth}
    \includegraphics[width=\textwidth]{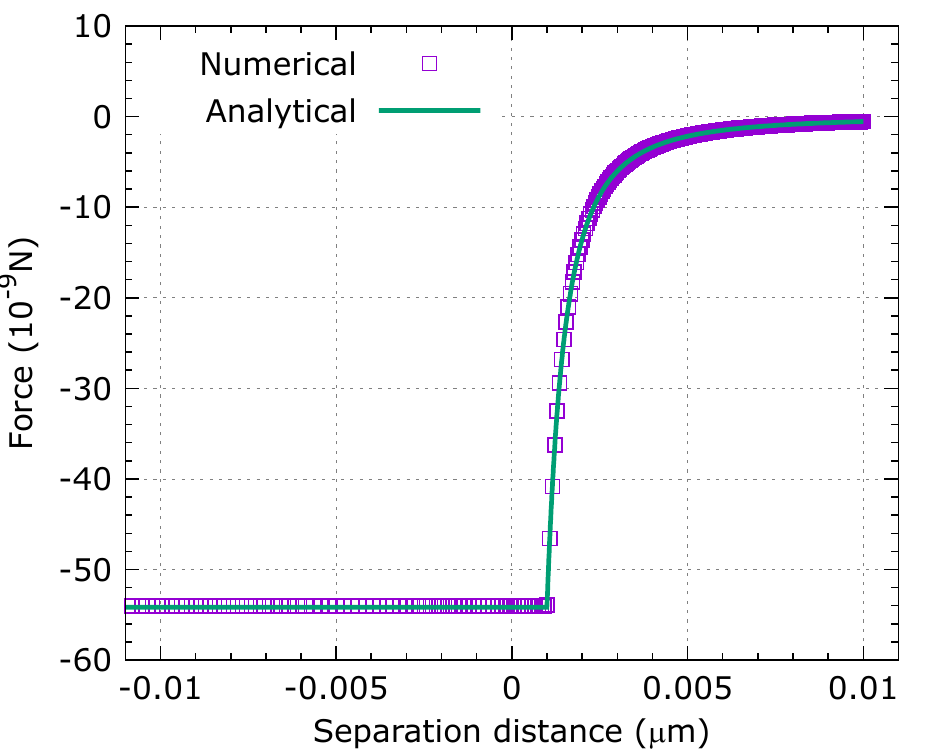}
    \caption{}
    \label{vdw_validationP2W}
  \end{subfigure}
  \caption{Validations of the Van der Waals force model: (a) particle to particle, (b) particle to wall.}
\label{vdw_validation}
\end{figure}
As shown in Figure~\ref{vdw_validation}, the purple dots are collected from the numerical simulations whereas the solid lines represent the corresponding analytical solutions. It can be seen from the two figures are that the analytical solution approximates zero when the separation distance becomes larger, and the force maintains a constant value when the separation distance is smaller than the cutoff distance $h_{\text{min}}$. For both Van der Waals interactions between two particles and a particle with a wall, the newly implemented model shows perfect agreement with the results given by the analytical solutions in Eqs.~\ref{VdWP2P} and \ref{VdWP2W}.

\subsection*{Voronoi tessellation for granular media: calculations of the local packing fraction with Voro++}
\label{voro}
Voronoi tessellation is used to calculate the local packing properties, e.g. the local packing fraction and packing structures of either mono-disperse or poly-disperse granular systems in this work. 
\begin{figure}[h]
  \begin{center}
    \includegraphics[width=0.35\textwidth]{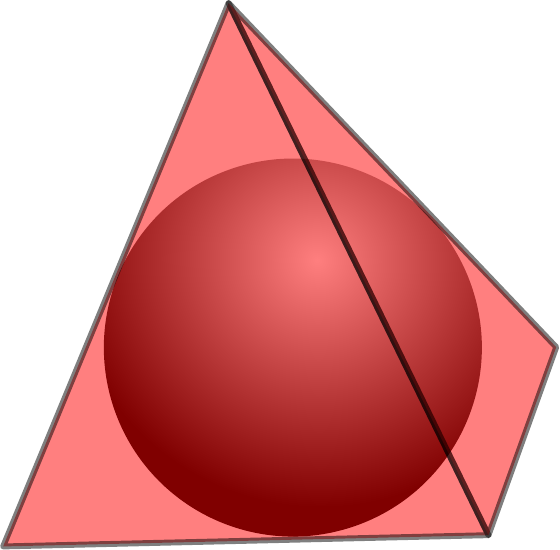}
  \end{center}
  \caption{Schematic diagram of a sphere enclosed by its voronoi cell.}
  \label{voronoiCell}
\end{figure}
The local packing fraction $\phi_f$ is defined by 
\begin{equation}
\phi_f = \frac{V_s}{V_c},
\end{equation}
where $V_s$ and $V_c$ are the volume of the sphere and its voronoi cell, namely the tetrahedron as shown in Figure~\ref{voronoiCell}, respectively. The open-source Voronoi tessellation code Voro++ is used to generate voronoi cells and calculate the volume of each voronoi cell.

\begin{figure}[h]
  \begin{center}
    \includegraphics[width=0.85\textwidth]{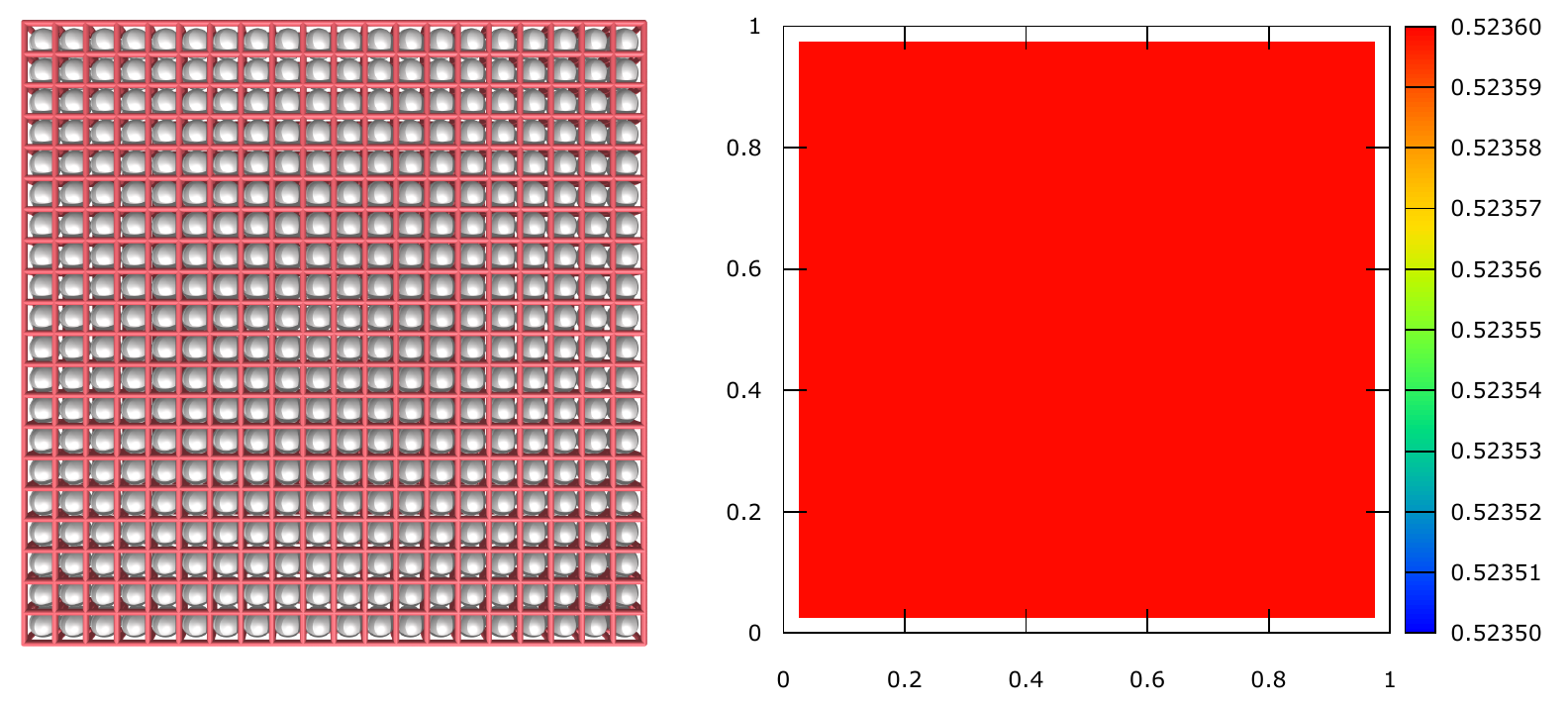}
  \end{center}
  \caption{(a) voronoi cells of the SCP, (b) local packing fraction of SCP.}
  \label{voroSCP}
\end{figure}

Two different cases are presented to demonstrate the performance of the Voro++ code in calculating the local packing fractions for the Simple Cubic Packing (SCP) and Hexagonal Close Packing (HCP). A single layer of SCP is shown in Figure~\ref{voroSCP}, 
for which the analytical solution of the packing fraction is given by
\begin{equation}
{\phi_f}^{\text{SCP}} = \frac{V_p}{V_c}=\frac{\frac{4 \pi R^3}{3}}{(2R)^3}=\frac{\pi}{6}= 0.5236.
\end{equation}
The image on the right-hand side of Figure~\ref{voroSCP} is the color map of the local packing fraction of the SCP. It proves that the local packing fraction calculated with Voro++ agrees well with the corresponding analytical solution.

\begin{figure}[h]
  \begin{center}
    \includegraphics[width=0.85\textwidth]{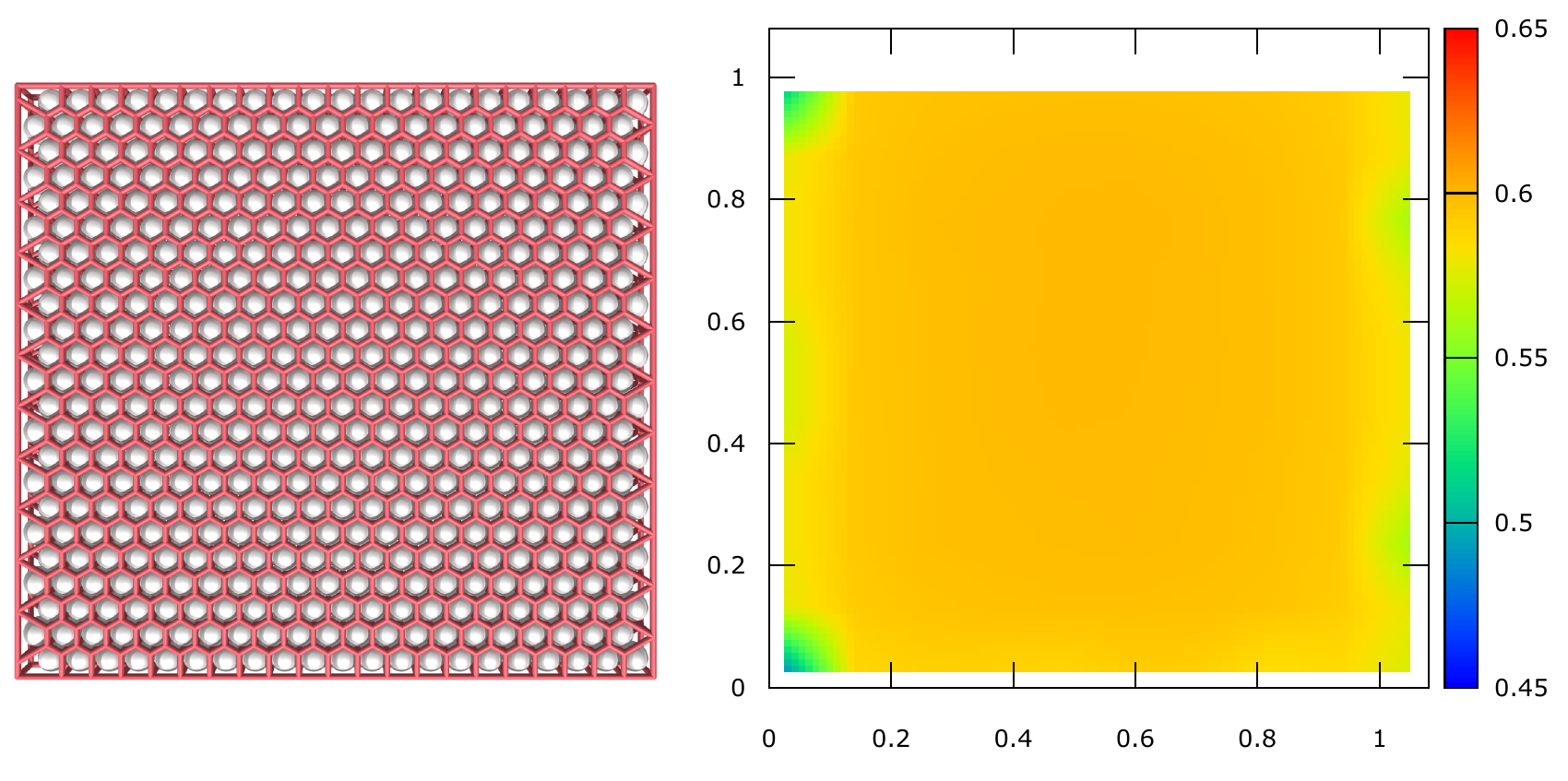}
  \end{center}
  \caption{(a) voronoi cells of the HCP, (b) local packing fraction of HCP.}
  \label{voroHCP}
\end{figure}

Similarly, a single layer of HCP is shown in Figure~\ref{voroHCP}, where a periodic boundary condition is applied along the horizontal direction. 
The analytical solution for the single layer HCP except for the local packing fraction on the boundary is given by
\begin{equation}
{\phi_f}^{\text{HCP}} = \frac{V_p}{V_c}=\frac{\frac{4 \pi R^3}{3}}{\frac{3 \sqrt{3}(\frac{2R}{\sqrt{3}})^22R}{2}}=\frac{\pi}{3 \sqrt{3}}= 0.6046.
\end{equation}
The local packing fraction calculated by Voro++ is shown in the right-hand side of Figure~\ref{voroHCP}, where a good agreement is found between the local packing fraction calculated by Voro++ and the analytical solution.

\end{spacing}

\bibliographystyle{elsarticle-num}
\bibliography{Refs}

\end{document}